\documentclass[sigconf]{acmart}

\AtBeginDocument{%
  }

\usepackage{multirow}
\usepackage{tabularx}
\usepackage{tcolorbox}

\usepackage{booktabs}
\usepackage{graphicx}

\usepackage{subcaption}
\usepackage{enumitem}
\usepackage{makecell}

\begin{document}

\title{Code Fingerprints: Disentangled Attribution of LLM-Generated Code}

\author{
Jiaxun Guo \quad
Ziyuan Yang \quad
Mengyu Sun \quad
Hui Wang \quad
Jingfeng Lu \quad
Yi Zhang \\[4pt]
School of Cyber Science and Engineering, Sichuan University, Chengdu, China \\[4pt]
\texttt{
mtt88176@gmail.com,
cziyuanyang@gmail.com,
mysun2001999@163.com,
whmzfc@gmail.com,
jingfeng.lu@scu.edu.cn,
yzhang@scu.edu.cn
}
}

\renewcommand{\shortauthors}{Trovato et al.}

\begin{abstract}
The rapid adoption of Large Language Models (LLMs) has transformed modern software development by enabling automated code generation at scale. While these systems improve productivity, they introduce new challenges for software governance, accountability, and compliance. Existing research primarily focuses on distinguishing machine-generated code from human-written code; however, many practical scenarios—such as vulnerability triage, incident investigation, and licensing audits—require identifying which LLM produced a given code snippet. In this paper, we study the problem of model-level code attribution, which aims to determine the source LLM responsible for generated code. Although attribution is challenging, differences in training data, architectures, alignment strategies, and decoding mechanisms introduce model-dependent stylistic and structural variations that serve as generative fingerprints. Leveraging this observation, we propose the Disentangled Code Attribution Network (DCAN), which separates Source-Agnostic semantic information from Source-Specific stylistic representations. Through a contrastive learning objective, DCAN isolates discriminative model-dependent signals while preserving task semantics, enabling multi-class attribution across models and programming languages. To support systematic evaluation, we construct the first large-scale benchmark dataset comprising code generated by four widely used LLMs (DeepSeek, Claude, Qwen, and ChatGPT) across four programming languages (Python, Java, C, and Go). Experimental results demonstrate that DCAN achieves reliable attribution performance across diverse settings, highlighting the feasibility of model-level provenance analysis in software engineering contexts.\footnote{The dataset and implementation are publicly available at \url{https://github.com/mtt500/DCAN}.}
\end{abstract}


\begin{teaserfigure}
  \centering
  \includegraphics[width=\textwidth]{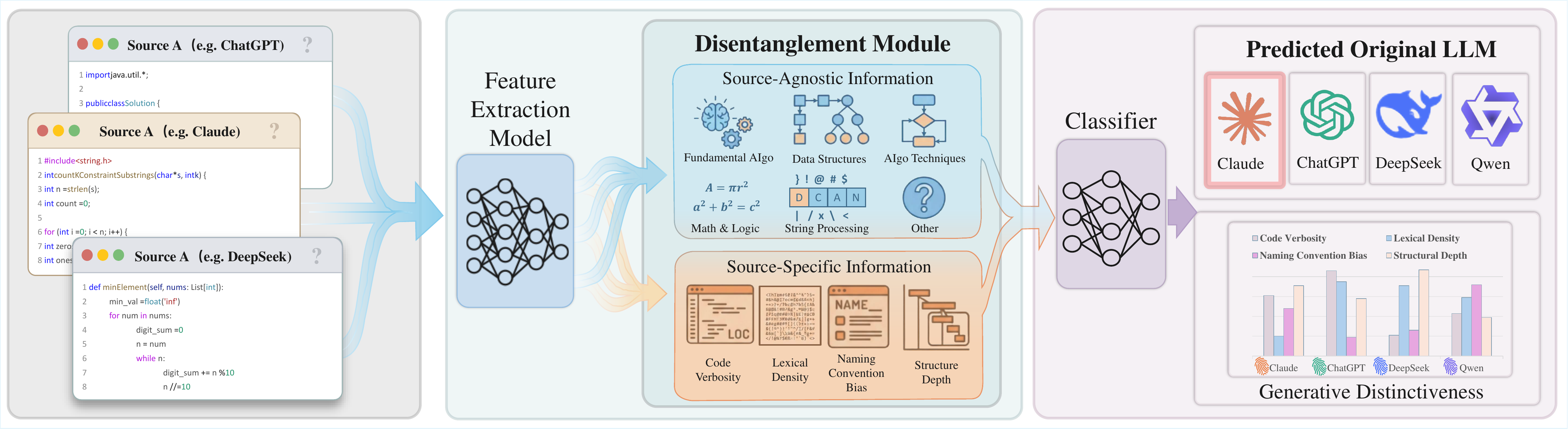}
  \caption{An overview of the proposed framework. Given code snippets generated by different LLMs, our approach disentangles the latent representation. An attribution classifier utilizes the extracted fingerprints to identify the source model.}
  \label{fig:overview}
\end{teaserfigure}

\received{20 February 2007}
\received[revised]{12 March 2009}
\received[accepted]{5 June 2009}

\maketitle

\section{Introduction}
Large Language Models (LLMs) have demonstrated remarkable capabilities in generating syntactically correct and functionally coherent code across multiple programming languages~\cite{liu2024deepseekv2, dubey2024llama, jiang2024survey}. These advances substantially lower the barrier to programming and improve developer productivity, which leads to their widespread adoption in real-world software development workflows~\cite{jimenez2023swe, poldrack2023ai}.
However, the increasing reliance on LLM-generated code introduces new challenges related to software provenance, security accountability, and intellectual property compliance. When vulnerabilities, malicious logic, or licensing conflicts are identified, determining the origin of generated artifacts becomes essential. Beyond distinguishing machine-generated code from human-written code, many practical software engineering scenarios require identifying which model produced a given code snippet. This motivates the problem of \textbf{\textit{LLM Code Source Attribution (LLMCSA)}}, which aims to attribute generated code to its originating LLM.

Research on code attribution builds upon prior efforts in detecting machine-generated natural language text~\cite{mitchell2023detectgpt, wu2025wrote, koike2024outfox}. More recently, authorship detection techniques have been extended to source code~\cite{orel2025codet}, which leverages pre-trained code foundation models such as CodeBERT~\cite{feng2020codebert} and UniXCoder~\cite{guo2022unixcoder}. While effective for binary detection settings, these methods are limited in realistic multi-source scenarios, as they do not explicitly model the fine-grained distinctions required to differentiate among multiple LLM providers.

LLMCSA is inherently challenging. For the same programming task, different LLMs often adopt similar solution strategies and follow fixed syntactic rules, which result in superficially similar outputs.
At the same time, LLMs differ in training corpora, architectural configurations, alignment strategies, and decoding mechanisms. These differences introduce variations in stylistic conventions, structural organization, and token-level preferences, which form implicit generative fingerprints that are difficult to capture with conventional detection methods.

We argue that effective LLMCSA requires disentangling two complementary types of information embedded in code representations: (1) \textbf{\textit{Source-Agnostic Information,}} which captures task-dependent functional semantics shared across models; and
(2) \textbf{\textit{Source-Specific Information,}} which encodes model-specific stylistic and structural fingerprints.
Existing studies on machine-generated code detection do not explicitly separate these factors, limiting their LLMCSA capability.

To address this challenge, we propose the Disentangled Code Attribution Network (DCAN), a novel LLMCSA framework that disentangles source-agnostic information from source-specific information. 
We align source-agnostic representations of code generated by different models for the same task, thereby isolating model-dependent components in the residual representation. This disentangled design enables more accurate and robust model-level attribution.

To systematically evaluate the performance, we construct the first large-scale benchmark LLMCSA dataset, comprising \textbf{91,804} code samples generated by four mainstream LLMs (DeepSeek ~\cite{liu2024deepseek}, Claude~\cite{anthropic2024claude3}, Qwen ~\cite{bai2023qwen}, and ChatGPT ~\cite{achiam2023gpt}), across \textbf{four} programming languages (C, Go, Java, and Python).
The dataset is collected under two coding settings (w/ and w/o comments) using a controlled generation and quality-control pipeline to ensure reliability and diversity.
Our contributions are summarized as follows:
\begin{itemize}
\item We introduce the LLMCSA task, a new software provenance problem that enables attribution of LLM-generated code to its originating provider.
\item We construct the first publicly available LLMCSA benchmark dataset with 91,804 samples across four programming languages and four LLMs under two coding settings.
\item We propose DCAN, a disentanglement-based attribution framework that explicitly separates source-agnostic information from source-specific information, enabling robust and accurate model-level attribution.
\end{itemize}

\begin{figure*}[!t]
  \centering
  \includegraphics[width=.9\textwidth]{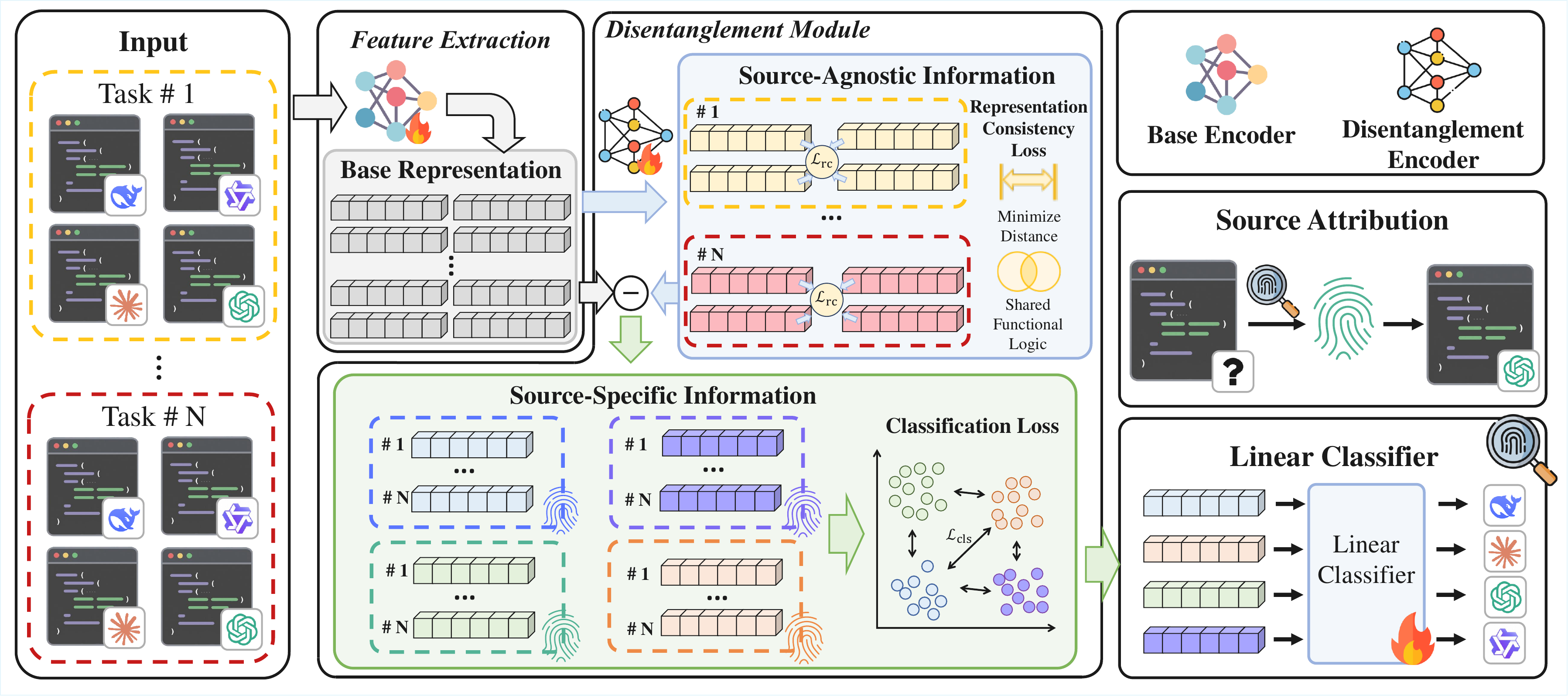}
  \vspace{-10pt}
  \caption{Overview of the DCAN framework.}
  \vspace{-10pt}
  \label{fig:model}
\end{figure*}

\section{Related Work}

Existing works on code provenance analysis can be broadly categorized into two directions: distinguishing human-written code from machine-generated code, and multi-class source attribution.

\textbf{Human vs. Machine Code Discrimination.} Early approaches adapted statistical NLP metrics to detect AI-generated code~\cite{xu2024detecting}. However, subsequent analyses by Pan \textit{et al.}~\cite{pan2024assessing} and Yang \textit{et al.} \cite{yang2023zero} demonstrated that such methods are ineffective for source code due to its rigid syntax, low entropy, and strong structural constraints. 
To address these limitations, later works incorporated structural and semantic features. Xu \textit{et al.} \cite{xu2025distinguishing} proposed \textit{CodeGPTSensor}, which employs contrastive learning on ``original vs. rewritten'' code pairs, leveraging the observation that LLMs tend to rewrite generated code with limited stylistic variation~\cite{ye2025uncovering}. 
From an explainability perspective, Bulla \textit{et al.} \cite{Bulla2024excode} introduced \textit{EX-CODE} to detect AI-generated code while highlighting interpretable indicators such as redundant imports and characteristic variable naming patterns.

\textbf{Model Attribution.} Identifying the specific generating model is significantly more challenging than binary detection. Existing attribution methods can be divided into active and passive approaches. Active methods rely on watermarking mechanisms embedded during generation. For example, Lu \textit{et al.} \cite{lu2025source} proposed \textit{WASA}, which embeds Unicode patterns into generated outputs, and Suresh \textit{et al.} \cite{suresh2024watermarking} introduced an AST-level watermarking scheme that inserts syntactically redundant structures. Although effective under controlled settings, these approaches require access to or modification of the generation process, limiting their applicability in real-world forensic scenarios.
Passive attribution operates solely on observed code without access to the generation pipeline and remains relatively underexplored. Recent works have begun to investigate stylistic patterns in specific programming languages~\cite{tihanyi2025hidden}. 
Bisztray \textit{et al.} \cite{bisztray2025know} introduced LLM-AuthorBench and applied supervised learning on handcrafted stylometric features to study model-level distinctions. 
Similarly, Huang \textit{et al.} \cite{huang2024can} utilized linguistic prompting to guide LLMs in extracting syntactic and vocabulary-based indicators for attribution. 
These studies provide initial evidence that LLMs possess distinct ``coding personalities''. However, their reliance on manually designed or language-specific features may limit scalability and cross-language generalization. 
In contrast, our work formulates model-level code attribution as a representation disentanglement problem. By explicitly disentangling source-agnostic information from source-specific information, our approach enables automated feature extraction and robust multi-language attribution without requiring access to the generation process.



\section{Methodology}
\label{sec:method}

\subsection{Problem Formulation}
\label{sec:problem_formulation}
We first formally define the LLMCSA task. Let $\mathcal{M} = \{M_1, M_2, \dots, M_K\}$ denote a finite set of $K$ candidate LLMs, where each $M_k$, for $k \in \{1, \dots, K\}$, represents a potential source model.
Let $\mathcal{T} = {t_1, t_2, \dots, t_N}$ denote a collection of programming tasks. Each task $t_n$ is defined by a natural language description and a set of functional requirements, such as input–output specifications.
We formulate a code generation scenario where, for each task $t \in \mathcal{T}$, a code snippet $c$ is generated conditioned on its specification. Consequently, we construct a dataset $\mathcal{D} = \{(c_{i,j}, y_i)\}_{i=1}^{L}$, where $L$ is the total number of samples. 
Each code snippet $c_{i,j}$ is generated to solve the task $t_{j}$, and $y_i \in {1, \dots, K}$ represents its ground-truth source model label.

The goal of LLMCSA is to learn a mapping function $f: \mathcal{C} \rightarrow \mathcal{Y}$, where $\mathcal{C}$ represents the code space and $\mathcal{Y} = \{1, \dots, K\}$ is the label space corresponding to the candidate models. In practice, this is formulated as a multi-class classification problem optimized via empirical risk minimization:
\begin{equation}
    \theta^* = \mathop{\arg\min}_{\theta} \frac{1}{L} \sum_{i=1}^{L} \mathcal{L}_{CE}(f_\theta(c_{i,j}), y_i),
\end{equation}
where $f_\theta$ denotes the attribution model parameterized by $\theta$, and $\mathcal{L}_{CE}$ represents the cross-entropy loss.


\noindent\textbf{Latent Disentanglement Hypothesis.} 
A fundamental challenge in LLMCSA lies in the fact that an observed code snippet $c$ is an entangled manifestation of two latent factors: \textit{Source-Agnostic Information}, dictated by the algorithmic task, and \textit{Source-Specific Information} inherent to the LLM. Conventional detection methods often learn representations dominated by task-dependent semantics, since functional correctness and solution structure contribute the prominent patterns in code. As a result, subtle model-specific fingerprints are easily overshadowed, leading to suboptimal LLMCSA performance.

To address this, we assume that the latent representation of a code snippet can be factorized into an \textit{additive composition} of two feature components: $h = E(c) = \mathbf{z}_c + \mathbf{z}_s$, where $E(\cdot)$ is an encoder, $\mathbf{z}_c$ denotes the source-agnostic information component, and $\mathbf{z}_s$ denotes the source-specific information component.


We assume approximate conditional invariance properties:
\begin{equation}
P(\mathbf{z}_c|t, y) \approx P(\mathbf{z}_c|t, y') \text{and} P(\mathbf{z}_s|t, y) \approx P(\mathbf{z}_s|t', y).
\end{equation}
That is, the source-agnostic information component $\mathbf{z}_c$ is primarily determined by the task and remains invariant across models, while the source-specific information component $\mathbf{z}_s$ depends on the source model and remains relatively stable across tasks.

Under this formulation, an ideal attribution function should rely primarily on $\mathbf{z}_s$ while remaining insensitive to $\mathbf{z}_c$. Our proposed framework explicitly enforces this disentanglement through tailored optimization objectives.

\subsection{Overview}
In this paper, we propose DCAN, a novel LLMCSA framework, which improving attribution performance by learning disentangled representations that separate \textit{source-agnostic} and \textit{source-specific} Information. An overview of DCAN is illustrated in Figure~\ref{fig:model}. DCAN first adopts an encoder to map each code snippet into a latent representation, which is subsequently decomposed to extract the source-specific information component through a disentanglement module. Then, LLMCSA is performed exclusively on the source-specific information component to ensure invariance to task-level semantics. To facilitate effective disentanglement, we introduce a tailored training loss that explicitly ensures the separation between source-agnostic and source-specific information components.


\subsection{Feature Extraction}
\label{subsec:representation_learning}

At first, we employ the pretrained encoder UniXcoder~\cite{guo2022unixcoder} to obtain latent representations for each code snippet. UniXcoder jointly models source code together with auxiliary structures such as comments and abstract syntax trees, which enable structurally informed embeddings that capture both syntactic and semantic information.

Given a code snippet $c = \{x_1, x_2, \dots, x_R\}$, where $x_r$ denotes the $r$-th token in the tokenized sequence, we prepend a classification token \texttt{[CLS]} and feed the sequence into the encoder. 
The Transformer layers model long-range dependencies through self-attention and produce contextualized hidden states for all tokens. Then, we extract the hidden state corresponding to the \texttt{[CLS]} token as follows:
\begin{equation}
h_{\text{base}} = \big[\text{Encoder}(\{x_{\texttt{[CLS]}}, x_1, x_2, \dots, x_R\})\big]_0 \in \mathbb{R}^{d},
\end{equation}
where $d$ is the hidden dimension, which is set to 768. $[\cdot]_0$ represents the hidden state of the \texttt{[CLS]} token. At this stage, $h_{\text{base}}$ serves as a unified latent representation that entangles both \textit{source-agnostic information} (i.e., task-dependent semantics) and the \textit{source-specific information} (i.e., model-specific generative fingerprints). 

\subsection{Feature Disentanglement}
\label{subsec:disentanglement}
The core component of DCAN is the \textit{Disentanglement Module}, which separates the \textit{source-specific} information from $h_{\text{base}}$. Since all candidate LLMs generate solutions for the same set of algorithmic tasks, task-dependent semantic features are shared across code samples generated by different LLMs.
By contrasting code samples generated by different LLMs for the same task, we can identify and align the shared task-level semantic features, thereby facilitating the separation of source-specific information from $h_{\text{base}}$.

To achieve this, we introduce a non-linear projection network to approximate the source-agnostic information $\mathbf{z}_c$. The projection process can be formulated as follows:
\begin{equation}
    h_{\text{com}} = \text{MLP}_{\text{com}}(h_{\text{base}}), 
\end{equation}
where $\text{MLP}_{\text{com}}$ consists of two linear layers interleaved with Batch Normalization and ReLU activation. The projection $h_{\text{com}}$ captures the task-dependent semantics shared across all candidate LLMs.

Next, we extract the source-specific information through a \textbf{subtractive decomposition}:
\begin{equation}
    h_{\text{spec}} = h_{\text{base}} - h_{\text{com}}.
\end{equation}

This design follows the assumption that $h_{\text{base}}$ admits an approximate additive decomposition  $h_{\text{base}} \approx h_{\text{com}} + h_{\text{spec}}$. 
From a representation learning perspective, if task semantics and model-dependent stylistic patterns affect the embedding in different ways, their contributions can be disentangled within the latent space. Thus, the embedding can be decomposed into two components: a task-dependent semantic component and a model-specific stylistic component.


By explicitly modeling and removing the source-agnostic information component $h_{\text{com}}$, the remaining representation $h_{\text{spec}}$ is encouraged to emphasize model-specific characteristics that are discriminative for LLMCSA.

\subsection{Optimization and Learning Objectives}
\label{subsec:optimization}

To ensure that the disentangled features are discriminative for LLMCSA, we propose a joint optimization strategy. The overall training objective is formulated as a weighted combination of two complementary loss functions: a \textbf{Source Classification Loss} applied to the source-specific information component, and a \textbf{Representation Consistency Loss} applied to the source-agnostic information component.

\subsubsection{Source Classification Loss}
First, to enforce the discriminative capability of the extracted source-specific information component, we attach a linear classifier to $h_{\text{spec}}$. The goal is to ensure that the model-specific fingerprints are preserved for accurate source identification. Given a batch of size $B$, we optimize the standard cross-entropy loss:
\begin{equation}
    \mathcal{L}_{cls} = - \frac{1}{B} \sum_{i=1}^{B} \log \left( P(y_i | h_{\text{spec}}^{i}) \right).
\end{equation}

The probability distribution over source models is computed using a linear classifier followed by a softmax function:
\begin{equation}
P(y \mid h_{\text{spec}}^{i})
=
\mathrm{Softmax}\!\left(W h_{\text{spec}}^{i} + b\right)_y,
\end{equation}
where $W \in \mathbb{R}^{K \times d}$ and $b \in \mathbb{R}^{K}$ are learnable parameters, and $(\cdot)_y$ denotes the predicted probability corresponding to class $y$. Minimizing $\mathcal{L}_{cls}$ encourages $h_{\text{spec}}$ to encode model-specific stylistic and structural fingerprints that facilitate accurate LLMCSA.

\subsubsection{Representation Consistency Loss}
For the decomposition to be meaningful, the learned projection $h_{\text{com}}$ must capture the task-dependent functional semantics shared across solutions generated by different LLMs for the same task.
If $h_{\text{com}}$ fails to encode this shared structure, subtracting it from $h_{\text{base}}$ would not effectively isolate model-specific fingerprints. To encourage task-level invariance, we introduce a Representation Consistency Loss $\mathcal{L}_{rc}$, which regularizes the common projection to be consistent across samples solving the same task.


Given a mini-batch of $B$ code samples $\mathcal{B} = \{c_{i,j}\}_{i=1}^{B}$, we group samples according to their task index $j$. We construct pairwise combinations of samples $(c_{i,j}, c_{q,j})$ such that they share the same task index $j$ and $i \neq q$.

We then minimize the cosine distance between their common representations:
\begin{equation}
\mathcal{L}_{rc}
=
\frac{1}{|\mathcal{S}|}
\sum_{(i,q,j)\in \mathcal{S}}
\big(1 - \cos(h_{\text{com}}^{i,j}, h_{\text{com}}^{q,j})\big),
\end{equation}
where $\mathcal{S} = \{ (i, q, j) \mid c_{i,j} \in \mathcal{B},\, c_{q,j} \in \mathcal{B},\, i \neq q \}$ denotes the set of all valid same-task sample triplets in the mini-batch.
This objective forces $h_{\text{com}}$ to encode task-dependent information. Consequently, when $h_{\text{com}}$ is subtracted from $h_{\text{base}}$, the residual component $h_{\text{spec}}$ is driven to emphasize model-specific information rather than shared functional semantics.

\subsubsection{Total Objective}
The final training objective integrates both attribution and disentanglement goals:
\begin{equation}
    \mathcal{L}_{total} = \mathcal{L}_{cls} + \lambda \mathcal{L}_{rc},
\end{equation}
where $\lambda$ is a hyperparameter that balances source discriminability and task-level semantic consistency. In our experiments, we set $\lambda = 0.2$ unless otherwise specified. By jointly minimizing this objective, DCAN captures model-specific fingerprints while reducing interference from shared functional semantics.

\section{Dataset Construction}

\subsection{Dataset Overview}
\begin{figure}[t]
  \centering
  \includegraphics[width=.9\columnwidth]{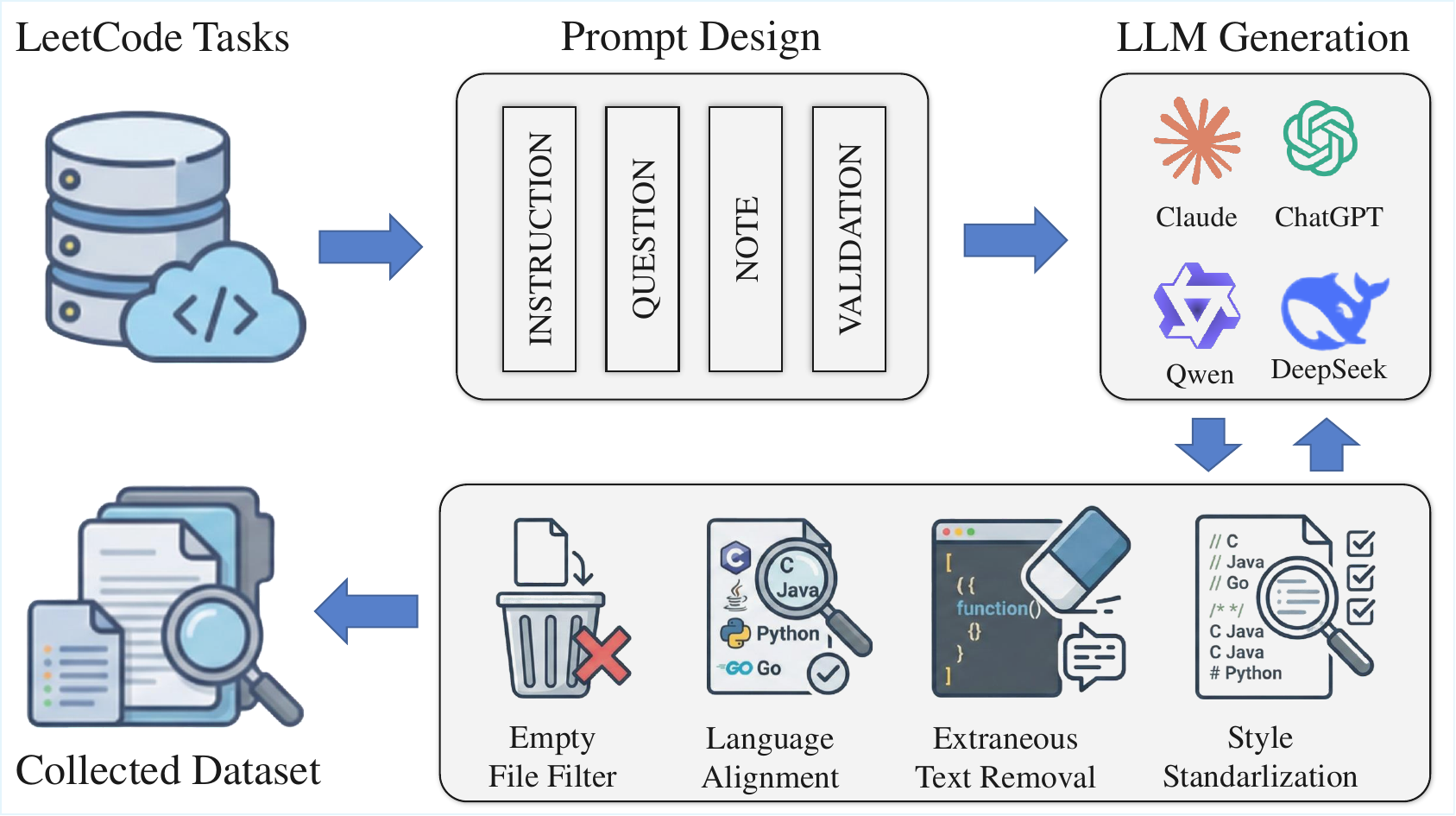}
  \vspace{-5pt}
  \caption{Dataset construction pipeline}
  \vspace{-15pt}
  \label{fig:dataset}
\end{figure}

To the best of our knowledge, LLMCSA has not been systematically studied in prior work, and there is currently no publicly available benchmark specifically designed for this task.
To facilitate reproducible evaluation and future research, we construct a large-scale LLMCSA dataset comprising 91,804 code samples. The dataset includes code generated by four widely used LLMs (DeepSeek, Claude, Qwen, and ChatGPT) across four programming languages (C, Go, Java, and Python).
To ensure comprehensive task coverage, we adopt the task set from LeetCodeDataset~\cite{xia2025leetcodedataset} as a unified benchmark. LeetCodeDataset contains 2,869 algorithmic tasks spanning diverse data structures and problem-solving strategies, providing broad functional variability.
For each task, we prompt every candidate LLM to generate corresponding solutions in all four programming languages. This design ensures balance across both the task dimension and the model dimension, preventing bias toward specific tasks or source models.
To further analyze attribution robustness under different code-generation scenarios, we explicitly consider two settings: (1) Plain Setting (w/o comments): models generate code without explanatory comments; and (2) Comment Setting (w/ comments): models generate code including inline or block comments. For each programming task, samples are generated under both settings, enabling controlled comparison of stylistic fingerprints with and without auxiliary natural-language content.
The overall dataset construction pipeline is illustrated in Figure~\ref{fig:dataset}.



\subsection{Construction Pipeline}
\subsubsection{Prompt Design}
For each task and each model, we construct a standardized prompt that explicitly specifies both the target programming language and the desired generation setting. To ensure fairness and consistency, the same set of prompts is used across all LLMs. Besides, the prompts are publicly released to promote reproducibility and facilitate a clearer understanding of our dataset.

\textbf{\textit{Plain Setting:}} In the plain setting, the prompt consists of three core components: \textit{INSTRUCTION}, \textit{QUESTION}, and \textit{NOTE}. The \textit{INSTRUCTION} component defines the model’s role and specifies the requirement to generate a correct program in the target language. The \textit{QUESTION} component provides the complete task specification sampled from the LeetCodeDataset. The \textit{NOTE} component imposes strict generation constraints, including the use of the designated programming language and the prohibition of comments.

\textbf{\textit{Comment Setting:}} The comment setting retains the same three-component structure but introduces modifications. 
Unlike the plain setting, the constraints require the inclusion of in-line comments to elucidate the program logic. However, prompting for explanations often leads models to produce additional conversational text (e.g., polite preambles like ``Here is the solution'' or explanatory postscripts) surrounding the code block. To suppress such artifacts, we incorporate a fourth component, \textsc{VALIDATION}.
This component acts as a self-correction mechanism, directing the model to retrospectively inspect its output and verify that the response consists solely of a code block containing the required comments, without any external natural language content. If extraneous conversational text is detected, the model is instructed to regenerate the output while strictly adhering to the code-only format.

\subsubsection{Code Generation}
Following the prompt design, we execute a controlled generation pipeline to construct the raw code corpus. 
We instantiate the standardized prompts with all 2,869 algorithmic tasks from the LeetCodeDataset and query four candidate LLMs to generate corresponding solutions. 
This process is conducted across all combinations of LLM, programming language, and coding setting. 
By strictly adhering to the predefined prompt protocols, we obtain a collection of initial model outputs, which served as the raw input for the subsequent data refinement phase.

\subsubsection{Data Refinement}

Despite carefully designed prompts, generated outputs may still violate formatting or content requirements. To ensure data quality, we design a dedicated refinement pipeline for the LLMCSA dataset, which consists of the following steps.

\textit{\textbf{Empty File Filter.}} We remove invalid outputs, including files that are empty or contain only whitespace characters. 

\textbf{\textit{Language Alignment.}} We perform language verification to ensure that each generated sample matches the designated programming language. Outputs that do not conform to the target language are discarded (e.g., Python-style syntax generated under Java, C, or Go settings).

\textbf{\textit{Narrative Text Removal.}} Raw LLM responses often contain natural-language explanations or introductory phrases (e.g., ``Here is the solution...''). 
We parse each response to extract only the content enclosed within code block delimiters (e.g., triple backticks), thereby removing extraneous textual descriptions. 

\textbf{\textit{Comment Standardization:}} 
We enforce comment consistency according to the designated generation mode. As comment syntax varies across programming languages, we apply language-specific pattern matching to identify comment segments while preserving string literals. 
For C, Java, and Go, we detect standard line comments (\texttt{//}) and block comments (\texttt{/* ... */}); for Python, we handle hash-based comments (\texttt{\#}) and docstrings.
In the Plain setting, all comments are removed to isolate algorithmic logic. In the Comment setting, comments are retained to preserve stylistic fingerprints that may contribute to source attribution.


\textbf{\textit{Adaptive Regeneration.}}
After the refinement steps above, low-quality samples are removed to ensure dataset reliability. To maintain both quality and completeness, we implement an iterative regeneration loop integrated with the refinement pipeline. Samples that fail any refinement step are discarded and flagged for regeneration. For these cases, we re-prompt the corresponding LLM to generate a new candidate solution. This regeneration cycle is repeated up to three times. If a valid solution cannot be obtained after the maximum number of retries, the corresponding solution is permanently excluded from the final dataset.

\subsection{Dataset Statistics}

Following the rigorous data generation and refinement pipeline, we obtain a high-quality LLMCSA benchmark dataset comprising \textbf{91,804} valid code samples. The dataset exhibits a strictly balanced distribution across two generation settings, \textit{Plain} and \textit{Comment}, each containing 45,902 samples. As described earlier, solutions for each task are generated by four different LLMs, and each LLM contributes approximately 11,476 samples per setting.
To establish a standardized benchmark for future research, we partition the dataset into training and testing sets. The training set comprises 84,508 samples derived from 2,641 distinct tasks, while the test set consists of 7,296 samples obtained from 228 reserved tasks.

\begin{figure*}[t]
    \centering
    \begin{subfigure}{0.33\textwidth}
        \centering
        \includegraphics[width=\columnwidth]{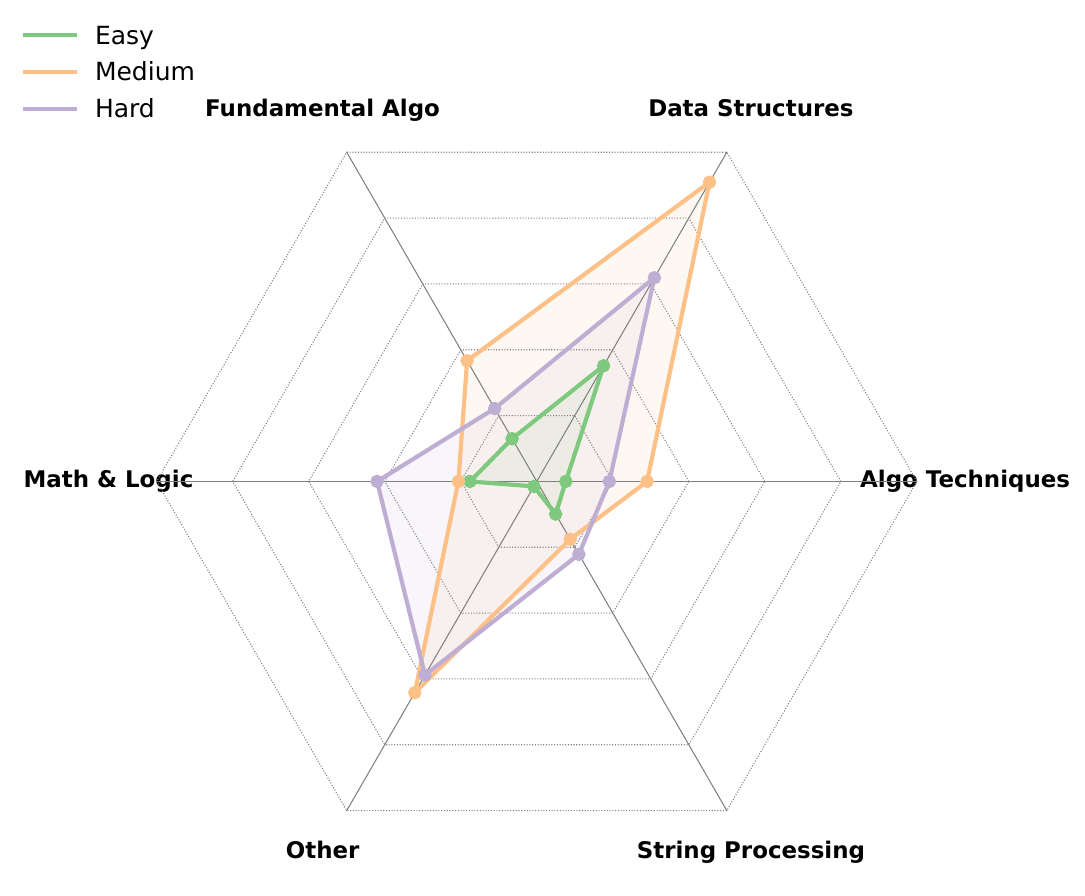}
        \caption{}
        \label{fig:stats_radar}
    \end{subfigure}
    \hfill
    \begin{subfigure}{0.33\textwidth}
        \centering
        \includegraphics[width=\columnwidth]{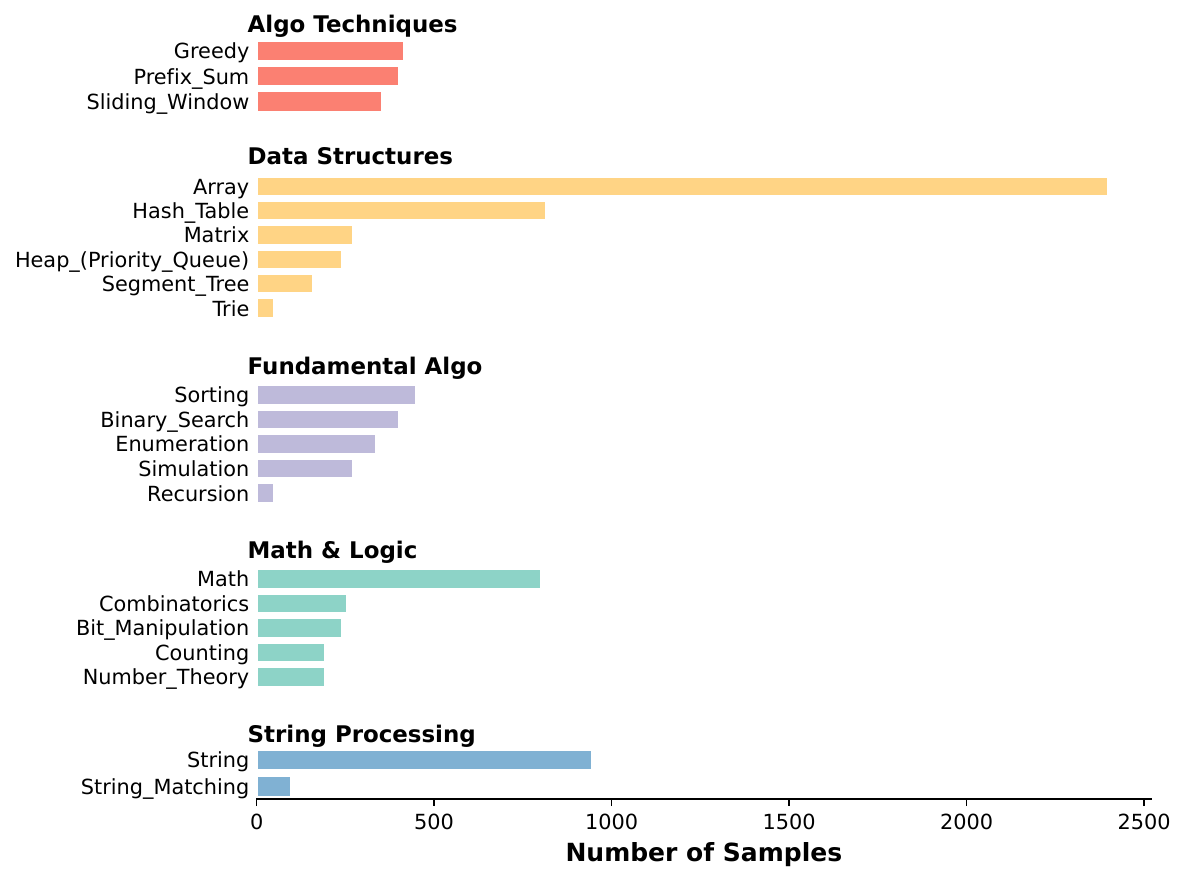}
        \caption{}
        \label{fig:stats_topic}
    \end{subfigure}
    \hfill
    \begin{subfigure}{0.33\textwidth}
        \centering
        \includegraphics[width=\columnwidth]{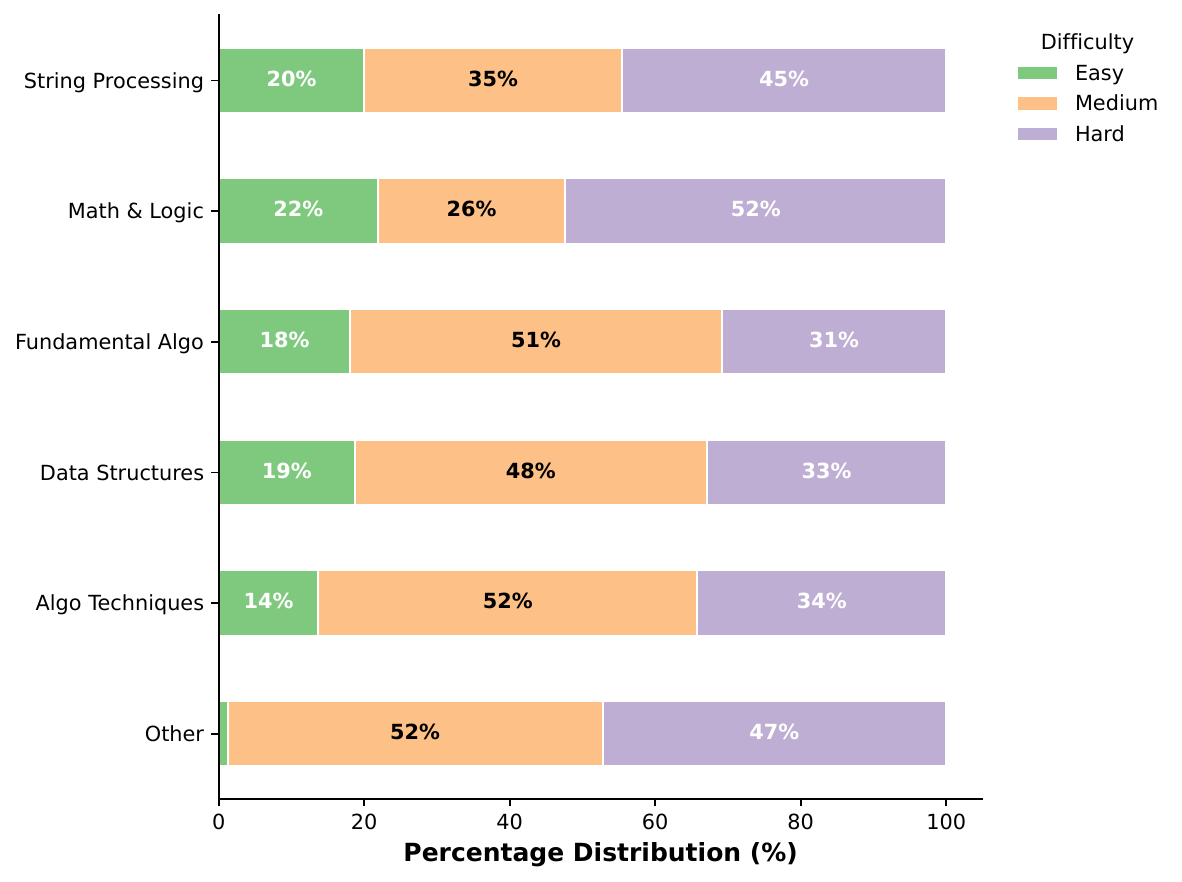}
        \caption{}
        \label{fig:stats_composition}
    \end{subfigure}
    
    \vspace{-5pt}
    \caption{\textbf{Dataset Diversity and Complexity Analysis.} 
    (a) \textbf{Difficulty Landscape:} Visualizes the distribution of task difficulties across major algorithmic domains. 
    (b) \textbf{Fine-grained Topic Distribution:} Provides a breakdown of sample counts across specific algorithmic tags. 
    (c) \textbf{Category Composition:} Illustrates the ratio of Easy, Medium, and Hard tasks within each category.}
    \label{fig:dataset_statistics}
    \vspace{-5pt}
\end{figure*}

To ensure comprehensive coverage and algorithmic diversity, the dataset spans five primary algorithmic domains: \textit{Data Structures}, \textit{Math \& Logic}, \textit{Fundamental Algorithms}, \textit{Algorithmic Techniques}, and \textit{String Processing}. The corresponding domain-level statistics are shown in Figure~\ref{fig:dataset_statistics}(a). At a fine-grained granularity (Figure~\ref{fig:dataset_statistics}(b)), the benchmark includes 21 algorithmic tags, ranging from fundamental topics like \textit{Arrays} and \textit{Sorting} to more complex optimization problems such as \textit{Dynamic Programming}. Furthermore, the dataset maintains a balanced difficulty distribution (Figure~\ref{fig:dataset_statistics}~(a) and~\ref{fig:dataset_statistics}~(c)), which enables systematic evaluation of attribution performance across varying levels of programming complexity. 

\section{Experiment Design}

\subsection{Research Questions}
\label{sec:Research Questions}

To systematically evaluate the effectiveness and robustness of the proposed DCAN framework, we investigate the following four research questions~(RQs):


    \noindent \textit{\textbf{RQ1. Generative Distinctiveness.}} \textit{Do different LLMs exhibit distinguishable generative preferences and distributional differences that can serve as attribution signals?}

    \noindent \textit{\textbf{RQ2. Attribution Feasibility.}} \textit{Is accurate model-level attribution of LLM-generated code achievable across multiple programming languages?}
    
    \noindent \textit{\textbf{RQ3. Mechanism Validity.}} \textit{Does the proposed disentanglement module improve attribution performance by isolating Source-Specific information from shared task semantics?}

    \noindent \textit{\textbf{RQ4. Robustness and Generalization.}} \textit{How robust is DCAN under varying data conditions and evaluation settings?}

\subsection{Experimental Settings}

\noindent \textbf{Baselines.} As no previous methods are specifically designed for the LLMCSA task, we adapt two advanced models, GPTSniffer~\cite{nguyen2024gptsniffer} and CodeGPTSensor~\cite{xu2025distinguishing}, as baselines. These models are originally proposed for distinguishing human-written code and machine-generated code and are extended here to the multi-class source attribution setting.

\noindent \textbf{Metrics.} We evaluate all methods using standard multi-class classification metrics, including accuracy and F1-score.

\noindent \textbf{Implementation.}
We implement DCAN using PyTorch. The encoder is initialized from the pre-trained UniXcoder~\cite{guo2022unixcoder}. The model is trained with the AdamW optimizer using a learning rate of $5\times10^{-5}$, a weight decay of 0.01, and a batch size of 32. The maximum input sequence length is set to 512. Training is performed for 12 epochs with a linear learning-rate warmup over the first 500 steps. All experiments are conducted on a single NVIDIA A100 GPU.


\section{Experimental Results}
\label{sec:results}


\newtcolorbox{findingbox}{
    colback=gray!10, 
    colframe=black,  
    boxrule=0.8pt,   
    left=4pt, right=4pt, top=3.5pt, bottom=3.5pt, 
    fontupper=\small\bfseries 
}

\subsection{RQ1: Generative Distinctiveness}
\label{sec:stylometric_analysis}

To investigate whether distinguishable distributional differences exist in LLM-generated code, we conduct the experiments under both the \textit{Plain} and \textit{Comment} settings.
This section examines whether code generated by different LLMs exhibits consistent and measurable differences in syntactic structure and explanation style.

\begin{figure}[!t]
    \centering
    \begin{subfigure}{0.48\linewidth}
        \centering
        \includegraphics[width=\linewidth]{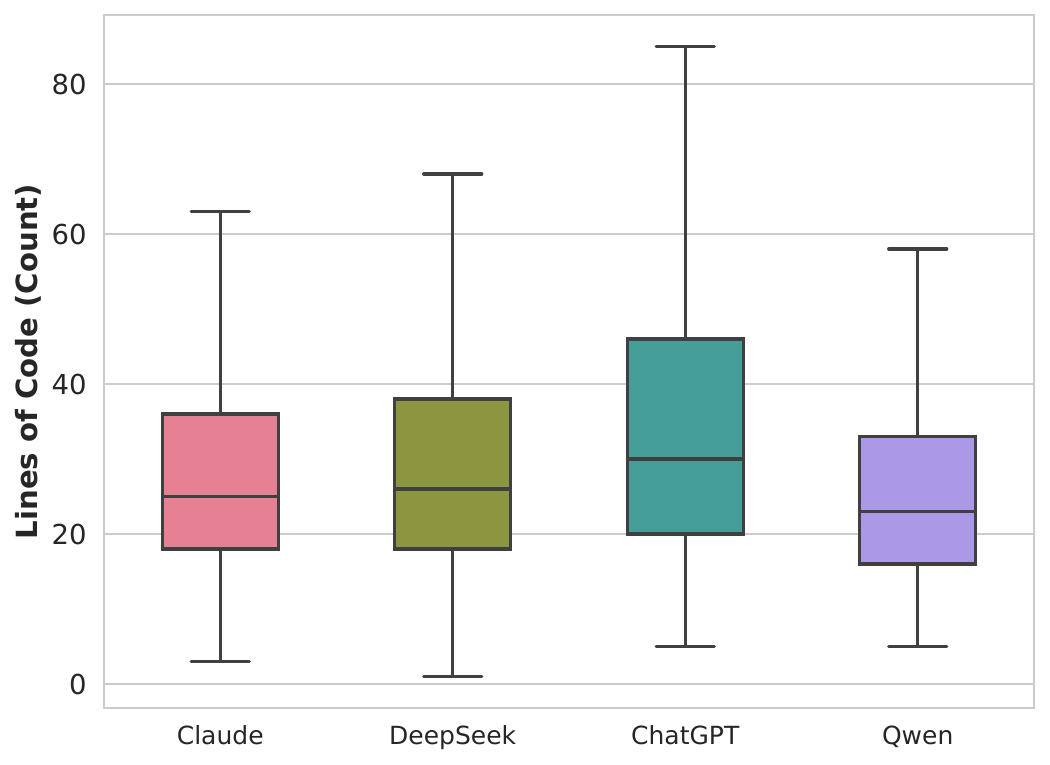}
        \caption{Code Verbosity (LOC)}
        \label{fig:loc}
    \end{subfigure}
    \hfill
    \begin{subfigure}{0.48\linewidth}
        \centering
        \includegraphics[width=\linewidth]{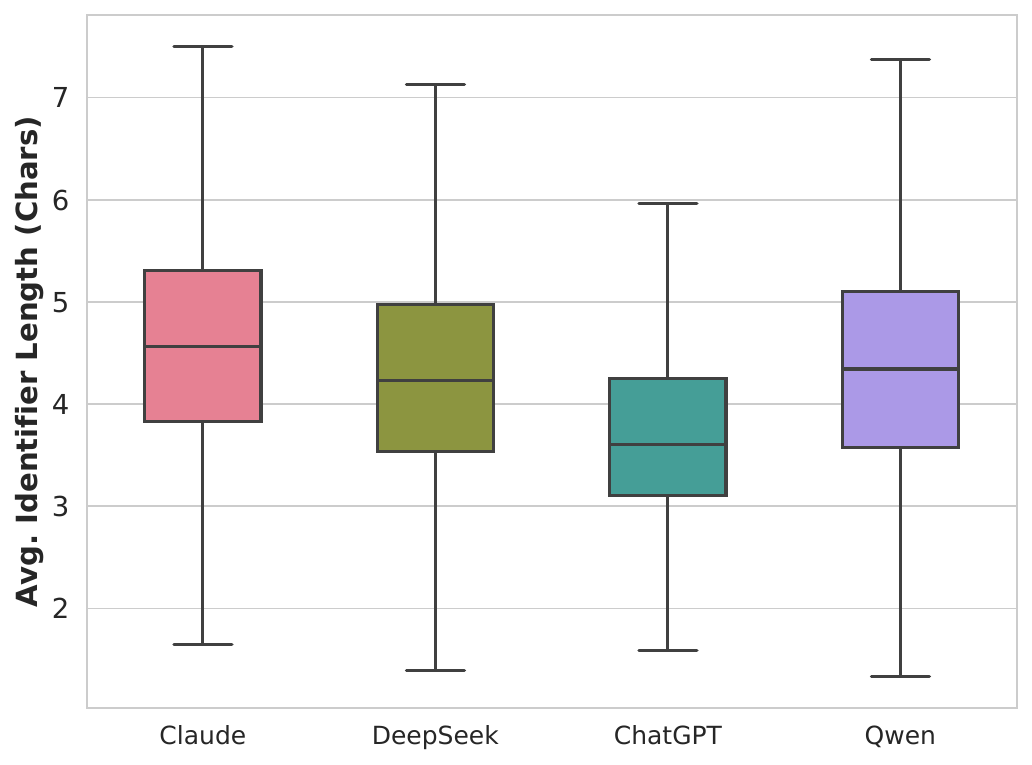}
        \caption{Lexical Density}
        \label{fig:id_len}
    \end{subfigure}
    
    \vspace{0.3cm} 
    
    \begin{subfigure}{0.48\linewidth}
        \centering
        \includegraphics[width=\linewidth]{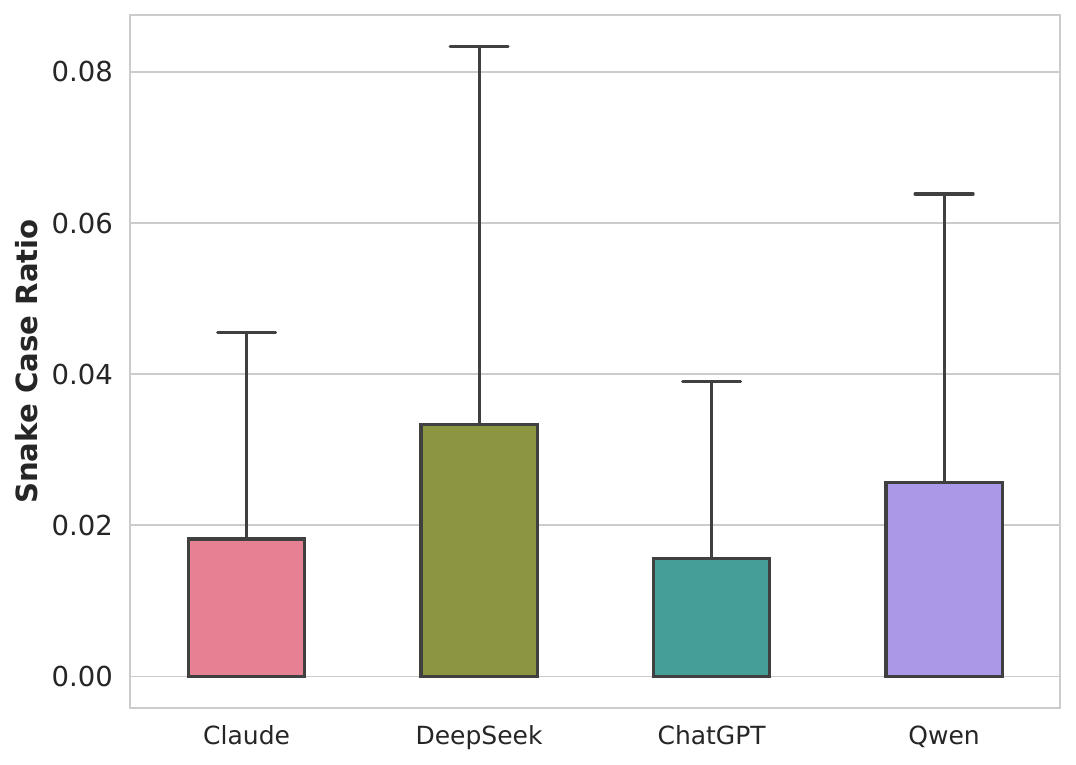}
        \caption{Naming Convention Bias}
        \label{fig:snake_case}
    \end{subfigure}
    \hfill
    \begin{subfigure}{0.48\linewidth}
        \centering
        \includegraphics[width=\linewidth]{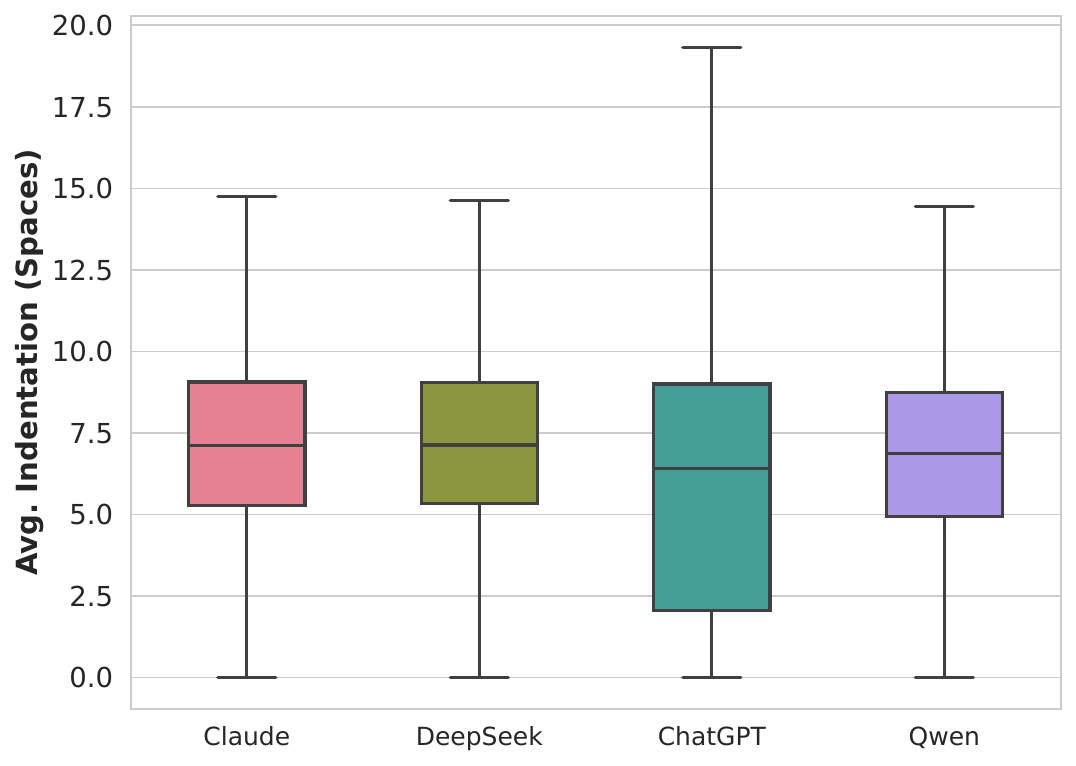}
        \caption{Structural Depth}
        \label{fig:indent}
    \end{subfigure}
    
    \caption{\textbf{Syntactic Distributional Differences Analysis.} The distinct medians and interquartile ranges reveal consistent generative personas across different LLMs.}
    \vspace{-10pt}
    \label{fig:syntax_boxplots}
\end{figure}

\textbf{Syntactic Distributional Differences in the Plain Setting.}
We first analyze stylistic distributional differences in the \textit{Plain} setting. To quantitatively capture syntactic distributional differences, we employed four metrics: 
(1) \textbf{Code Verbosity}, measured by total Lines of Code (LOC); 
(2) \textbf{Lexical Density}, defined as the average character length of identifiers; 
(3) \textbf{Naming Convention Bias}, quantified by the ratio of snake\_case variables (which deviate from Java's conventional camelCase style);
(4) \textbf{Structural Depth}, measured by the average indentation level. 

As illustrated in Figure~\ref{fig:syntax_boxplots}, the models demonstrate stable generation behaviors.
Regarding verbosity (Figure~\ref{fig:syntax_boxplots}(\subref{fig:loc})), ChatGPT exhibits the highest variance and median line count, which indicates a tendency toward more verbose implementations. In contrast, Qwen produces the most concise output with the lowest median LOC.
Lexical analysis (Figure~\ref{fig:syntax_boxplots}(\subref{fig:id_len})) shows that Claude favors longer identifiers, which suggests a preference for descriptive naming, whereas ChatGPT more frequently uses shorter variable names.
Figure~\ref{fig:syntax_boxplots}(\subref{fig:snake_case}) reveals different naming biases. Although Java conventionally favors camelCase, DeepSeek and Qwen exhibit a higher proportion of snake\_case identifiers compared to Claude and ChatGPT.
Finally, indentation statistics (Figure~\ref{fig:syntax_boxplots}(\subref{fig:indent})) indicate differences in structural consistency. Claude and DeepSeek maintain relatively stable indentation distributions, while ChatGPT exhibits greater variance.

\begin{figure}[!t]
    \centering
    \begin{subfigure}{0.48\columnwidth}
        \centering
        \includegraphics[width=\linewidth]{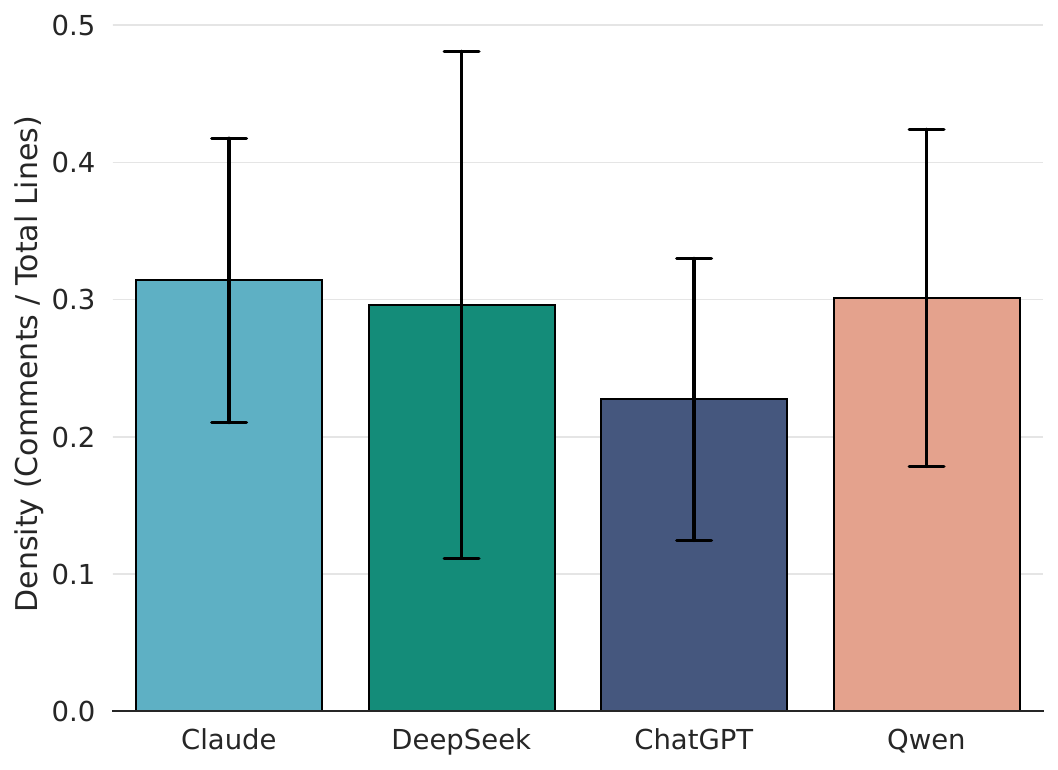}
        \caption{}
        \label{fig:comment_density}
    \end{subfigure}
    \hfill
    \begin{subfigure}{0.48\columnwidth}
        \centering
        \includegraphics[width=\linewidth]{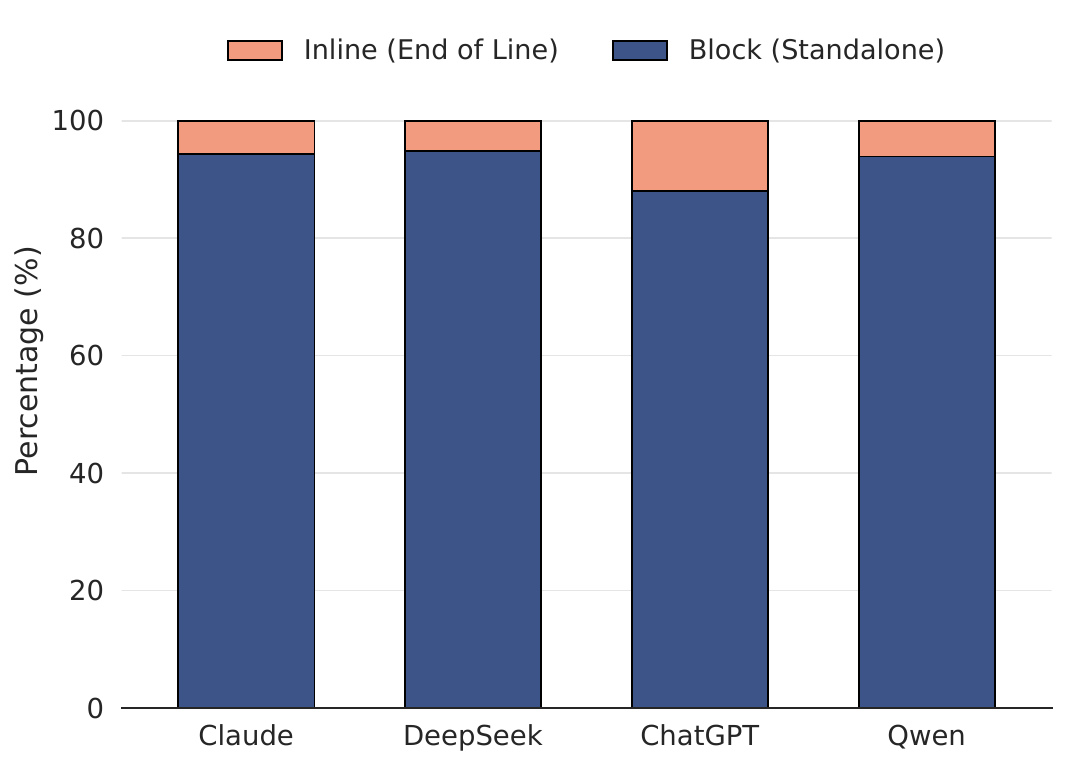}
        \caption{}
        \label{fig:stacked_comment_style}
    \end{subfigure}
    
    \vspace{-10pt}
    \caption{\textbf{Stylometric Analysis of Generated Comments.} (a) illustrates the varying degrees of helpfulness (density) across models, while (b) reveals distinct formatting preferences in annotation placement.}
    \label{fig:comment_analyze_combined}
    \vspace{-10pt}
\end{figure}

\textbf{Pedagogical and Formatting Distributional Differences in Comments.}
Beyond syntactic structure, natural-language comments serve as an additional modality for stylistic differentiation. As depicted in Figure~\ref{fig:comment_analyze_combined}, clear differences in explanatory behavior emerge. Figure~\ref{fig:comment_analyze_combined}(a) highlights variation in documentation volume. ChatGPT generates the lowest comment density, reflecting a preference for more self-contained logic with limited annotation.
Structural preferences further differentiate the models (Figure~\ref{fig:comment_analyze_combined}(b). While block comments (standalone comment lines) are generally dominant, ChatGPT displays a comparatively higher proportion of \textit{inline comments} (end-of-line annotations). Conversely, DeepSeek and Claude predominantly rely on block-level documentation, with minimal inline usage. These findings indicate that comment-level stylistic differences serve as reliable attribution signals.

\begin{findingbox}
    \textbf{Finding 1}: Different LLMs exhibit intrinsic and distinguishable generative distributional differences. Even when solving the same algorithmic task, models demonstrate consistent preferences in code structure, length, and commenting style.
\end{findingbox}

\subsection{RQ2: Attribution Feasibility}
\label{sec:rq1}

To evaluate the attribution capability of DCAN, we conduct a comprehensive evaluation across four programming languages under both the \textit{Plain} and \textit{Comment} settings. We first compare overall performance against the baselines. Subsequently, we perform a fine-grained analysis to assess robustness across different task domains and difficulty levels.

\subsubsection{Attribution Performance}
\label{sec:rq1_overall}
We evaluate DCAN against two adapted baselines under both generation settings. The comprehensive results are presented in Table~\ref{tab:main_result_combined}.

Under the \textit{Plain Setting}, DCAN consistently outperforms all baselines. While GPTSniffer achieves an average F1-score of 89.15\%, DCAN reaches 92.94\%. The improvements are consistent across all programming languages, demonstrating that DCAN achieves superior performance even without natural language information. 

Under the \textit{Comment Setting}, incorporating natural-language comments further improves attribution performance across all languages and source models. The average F1-score increases to 98.38\%. 
This improvement suggests that LLMs exhibit distinct linguistic fingerprints in explanatory text (e.g., phrasing choices and formatting conventions), which provide additional attribution cues.

\newcolumntype{Y}{>{\centering\arraybackslash}X}
\begin{table*}[t]
\centering
\caption{Performance comparison of LLM source attribution across four programming languages. \textbf{Panel A} reports results under the Plain setting, while \textbf{Panel B} reports results under the \textit{Comment} setting. The best results are highlighted in \textbf{bold}.}
\label{tab:main_result_combined}
\begin{tabular}
{c|cc|cc|cc|cc|cc}
\toprule
\multirow{2}{*}{\textbf{Method}} 
& \multicolumn{2}{c|}{\textbf{C}} 
& \multicolumn{2}{c|}{\textbf{Go}} 
& \multicolumn{2}{c|}{\textbf{Java}} 
& \multicolumn{2}{c|}{\textbf{Python}}
& \multicolumn{2}{c}{\textbf{Average}} \\
 & Acc. & F1 & Acc. & F1 & Acc. & F1 & Acc. & F1 & Acc. & F1 \\
\midrule
\multicolumn{11}{c}{\textit{\textbf{Panel A: Plain Setting}}} \\
\midrule
CodeGPTSensor 
& 82.46 & 82.48 
& 82.13 & 81.78 
& 76.43 & 75.89 
& 66.52 & 65.37 
& 76.89 & 76.38 \\

GPTSniffer 
& 92.17 & 92.17 
& 94.21 & 94.18 
& 91.04 & 91.01 
& 79.27 & 79.25 
& 89.17 & 89.15 \\

\textbf{DCAN (Ours)} 
& \textbf{96.16} & \textbf{96.16} 
& \textbf{96.28} & \textbf{96.27} 
& \textbf{93.17} & \textbf{93.20} 
& \textbf{85.28} & \textbf{86.13} 
& \textbf{92.72} & \textbf{92.94} \\

\midrule
\multicolumn{11}{c}{\textit{\textbf{Panel B: Comment Setting}}} \\
\midrule
CodeGPTSensor 
& 87.95 & 87.96 
& 97.78 & 97.78 
& 92.59 & 92.53 
& 90.68 & 90.61 
& 92.25 & 92.22 \\

GPTSniffer 
& 96.49 & 96.48 
& 96.98 & 97.01 
& 96.20 & 96.18 
& 93.46 & 93.48 
& 95.78 & 95.79 \\

\textbf{DCAN (Ours)} 
& \textbf{97.99} & \textbf{97.99} 
& \textbf{99.07} & \textbf{99.07} 
& \textbf{98.23} & \textbf{98.23} 
& \textbf{98.24} & \textbf{98.25} 
& \textbf{98.38} & \textbf{98.38} \\
\bottomrule
\end{tabular}
\vspace{-5pt}
\end{table*}

\subsubsection{Impact of Code Complexity and Domain}
\label{sec:rq1_fine_grained}

Beyond overall performance, we investigate whether DCAN exhibits bias toward specific task domains or complexity levels. We conduct a stratified evaluation across five primary algorithm categories and three difficulty levels. The results are summarized in Table~\ref{tab:fine_grained_performance}.

Interestingly, we observe a positive correlation between task difficulty and attribution performance. Unlike typical classification scenarios, where increased complexity often leads to performance degradation, DCAN frequently achieves higher F1-scores on Medium and Hard tasks than on Easy tasks. For instance, in the \textit{Algorithmic Techniques} category, the F1-score rises from 91.96\% on Easy tasks to 93.28\% on Hard tasks. One possible explanation is that Easy tasks often admit canonical and highly standardized solutions that appear nearly identical across LLMs, thereby reducing stylistic variability. In contrast, more complex tasks require intricate logic and implementation choices, which may amplify model-dependent fingerprints. These richer stylistic fingerprints appear to be effectively captured by DCAN.

Furthermore, DCAN demonstrates consistent performance across diverse task domains. Categories involving extensive syntactic variation, such as \textit{String Processing}, yield high attribution accuracy. Even in the \textit{Math \& Logic} category, where solutions are often concise and formulaic, DCAN maintains an F1-score of 90\%. These results indicate that the learned representations capture model-dependent fingerprints that persist across varying algorithm categories and difficulty levels.


\begin{table}[!t]
    \centering
    \caption{Performance metrics across different algorithm categories and difficulty levels.}
    \label{tab:fine_grained_performance}
    \begin{tabular}{llcccc}
        \toprule
        \textbf{Category} & \textbf{Difficulty} & \textbf{Prec.} & \textbf{Rec.} & \textbf{F1} & \textbf{Acc.} \\
        \midrule
        \multirow{3}{*}{\shortstack[l]{Algorithmic\\Techniques}} 
          & Easy   & 92.60 & 92.19 & 91.96 & 92.19 \\
          & Medium & 93.32 & 93.34 & 93.22 & 93.34 \\
          & Hard   & 93.74 & 93.43 & 93.28 & 93.43 \\
        \midrule
        \multirow{3}{*}{\shortstack[l]{Data\\Structures}} 
          & Easy   & 89.78 & 89.45 & 89.26 & 89.45 \\
          & Medium & 93.16 & 92.85 & 92.69 & 92.85 \\
          & Hard   & 91.67 & 91.57 & 91.41 & 91.57 \\
        \midrule
        \multirow{3}{*}{\shortstack[l]{Fundamental\\Algorithms}} 
          & Easy   & 91.28 & 90.63 & 90.57 & 90.63 \\
          & Medium & 92.15 & 92.16 & 92.06 & 92.16 \\
          & Hard   & 92.03 & 91.71 & 91.62 & 91.72 \\
        \midrule
        \multirow{3}{*}{\shortstack[l]{Math \&\\Logic}} 
          & Easy   & 86.54 & 85.42 & 85.17 & 85.42 \\
          & Medium & 90.21 & 90.01 & 89.76 & 90.01 \\
          & Hard   & 90.58 & 90.02 & 89.75 & 90.02 \\
        \midrule
        \multirow{3}{*}{\shortstack[l]{String\\Processing}} 
          & Easy   & 93.08 & 92.19 & 92.01 & 92.19 \\
          & Medium & 93.05 & 92.47 & 92.37 & 92.47 \\
          & Hard   & 91.81 & 91.47 & 91.21 & 91.47 \\
        \midrule
        \multirow{3}{*}{Other} 
          & Easy   & 84.58 & 84.38 & 83.83 & 84.38 \\
          & Medium & 93.87 & 93.08 & 92.89 & 93.08 \\
          & Hard   & 92.81 & 91.03 & 90.57 & 91.03 \\
        \bottomrule
    \end{tabular}
\end{table}

\begin{findingbox}
\textbf{Finding 2:} DCAN consistently achieves high attribution accuracy across different settings, demonstrating that model-level attribution is feasible across multiple programming languages.
\end{findingbox}

\subsection{RQ3: Mechanism Validity}

To evaluate the effectiveness of the proposed disentanglement mechanism, we conduct an ablation study on the Java Plain dataset, combining quantitative performance analysis and qualitative visual inspection.

\subsubsection{Quantitative Analysis}
We evaluate attribution performance using different features, including the raw feature $h_{\text{base}}$, the \textit{source-specific} information component $h_{\text{spec}}$, and the \textit{source-agnostic} information component $h_{\text{com}}$.


\begin{table}[t]
\centering
\caption{Ablation study on the Java Plain dataset.}
\label{tab:ablation_disentanglement}
    \begin{tabular}{llcc}
    \toprule
    \textbf{Feature Source} & \textbf{Symbol} & \textbf{Acc. } & \textbf{F1} \\
    \midrule
    Original Entangled & $h_{\text{base}}$ & 92.65 & 92.65 \\
    Source-Agnostic Only & $h_{\text{com}}$ & 24.89 & 09.96 \\
    \textbf{Source-Specific Only} & \boldmath$h_{\text{spec}}$ & \textbf{93.17} & \textbf{93.20} \\
    \bottomrule
    \end{tabular}%
\vspace{-10pt}
\end{table}

The results are presented in Table~\ref{tab:ablation_disentanglement}. 
The \textit{source-agnostic} information component $h_{\text{com}}$ achieves an accuracy of 24.89\%, which aligns with the 25\% random-guess baseline for four classes. This indicates that $h_{\text{com}}$ does not retain source-discriminative information.
In contrast, using only the \textit{source-specific} representation $h_{\text{spec}}$ achieves the highest performance (F1: 93.20\%), which indicates that the \textit{source-specific} information component captures the fingerprints necessary for attribution. These results suggest that DCAN effectively decouples the \textit{source-specific} information component $h_{\text{spec}}$ and the \textit{source-agnostic} information component $h_{\text{com}}$ from the raw representation.

\subsubsection{Visual Analysis of Representation Distribution}
\begin{figure}[t]
    \centering
    \begin{subfigure}{0.48\columnwidth}
        \includegraphics[width=\linewidth]{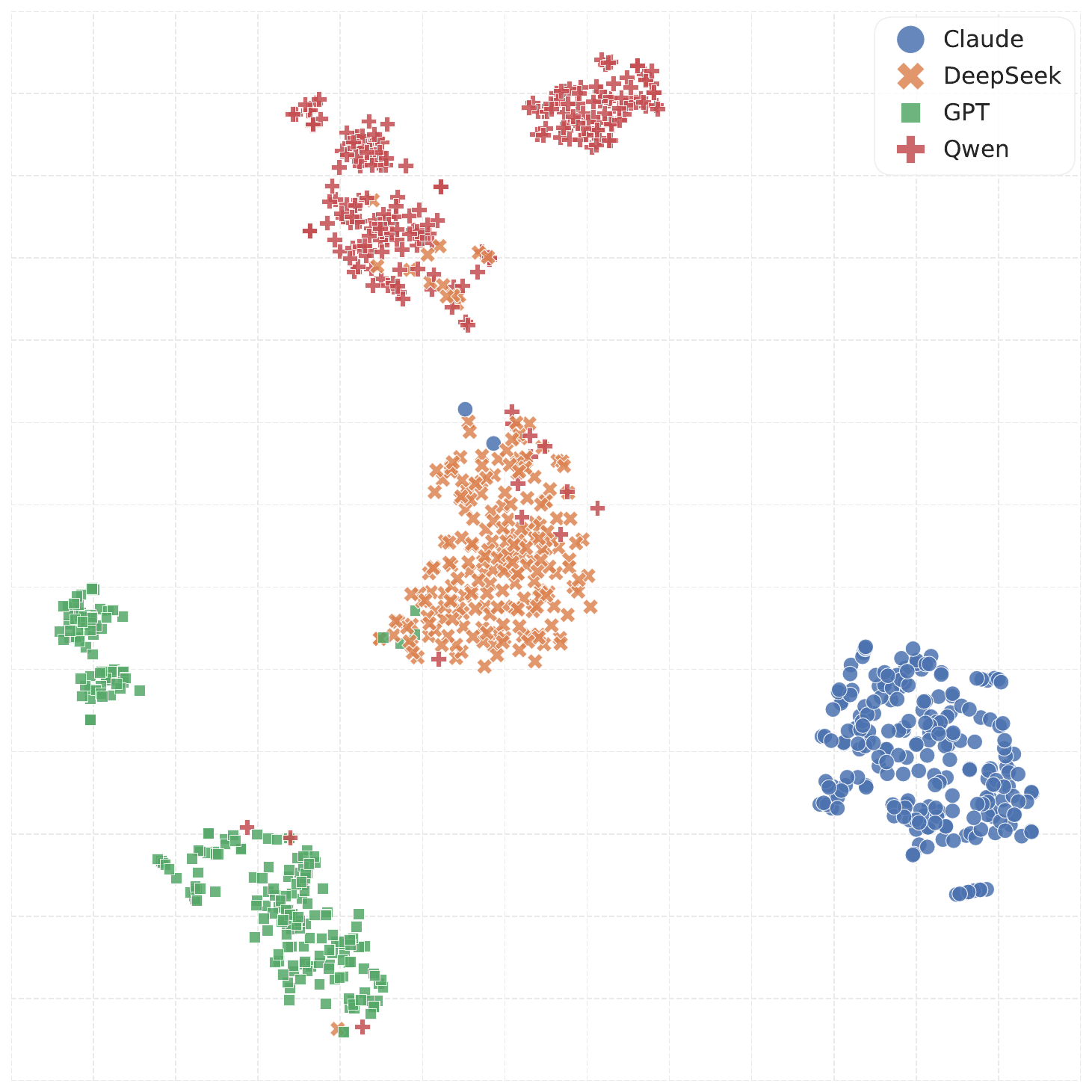}
        \caption{}
        \label{fig:go_plain_style_features}
    \end{subfigure}
    \hfill
    \begin{subfigure}{0.48\columnwidth}
        \includegraphics[width=\linewidth]{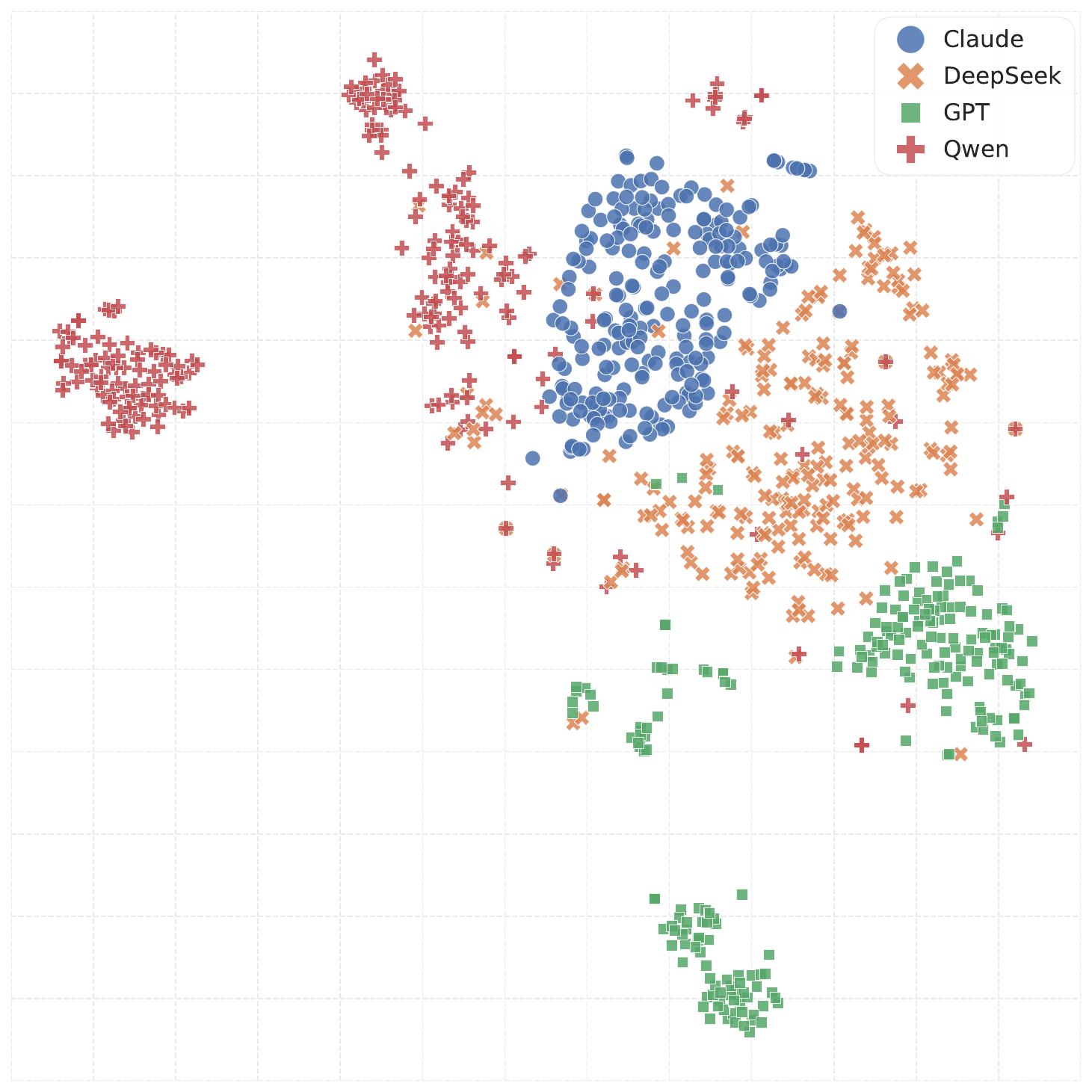}
        \caption{}
        \label{fig:go_plain_common_features}
    \end{subfigure}
    \vspace{-5pt}
    \caption{t-SNE visualizations of different features. (a) and (b) show the specific and common features, respectively.}
    \vspace{-10pt}
    \label{fig:tsne_feature}
\end{figure}

To further verify the effectiveness of our disentanglement strategy, we visualize the learned representations using t-SNE, as shown in Figure~\ref{fig:tsne_feature}. The comparison reveals a significant divergence in the distributions of the two feature spaces. 
In the \textbf{source-specific space} (Figure~\ref{fig:tsne_feature}(a)), samples form distinct and compact clusters corresponding to their source LLMs, which indicates that ${h}_{style}$ successfully captures the discriminative fingerprints required for attribution. In contrast, the \textbf{source-agnostic space} (Figure~\ref{fig:tsne_feature}(b)) exhibits substantial overlap among samples from different source models, with no clear decision boundaries. This comparison demonstrates that DCAN successfully isolates model-dependent fingerprints into the source-specific subspace, while reducing source-related information in the source-agnostic information component.


\begin{findingbox}
\textbf{Finding 3:} Explicitly separating the Source-Agnostic information component from the Source-Specific information component enhances attribution performance, indicating that isolating stylistic fingerprints from task-dependent semantics is critical for accurate model-level attribution.
\end{findingbox}

\subsection{RQ4: Robustness and Generalization}

\label{sec:lang_independence}

To comprehensively assess the robustness and generalizability of DCAN, we conduct experiments across three dimensions: (1) data efficiency across varying training scales, (2) the synergistic effect of unified multilingual training, and (3) zero-shot generalization to unseen programming languages.

\subsubsection{Data Efficiency Analysis}
\label{sec:data_efficiency}

\begin{figure}[t]
    \centering
    \includegraphics[width=.8\linewidth]{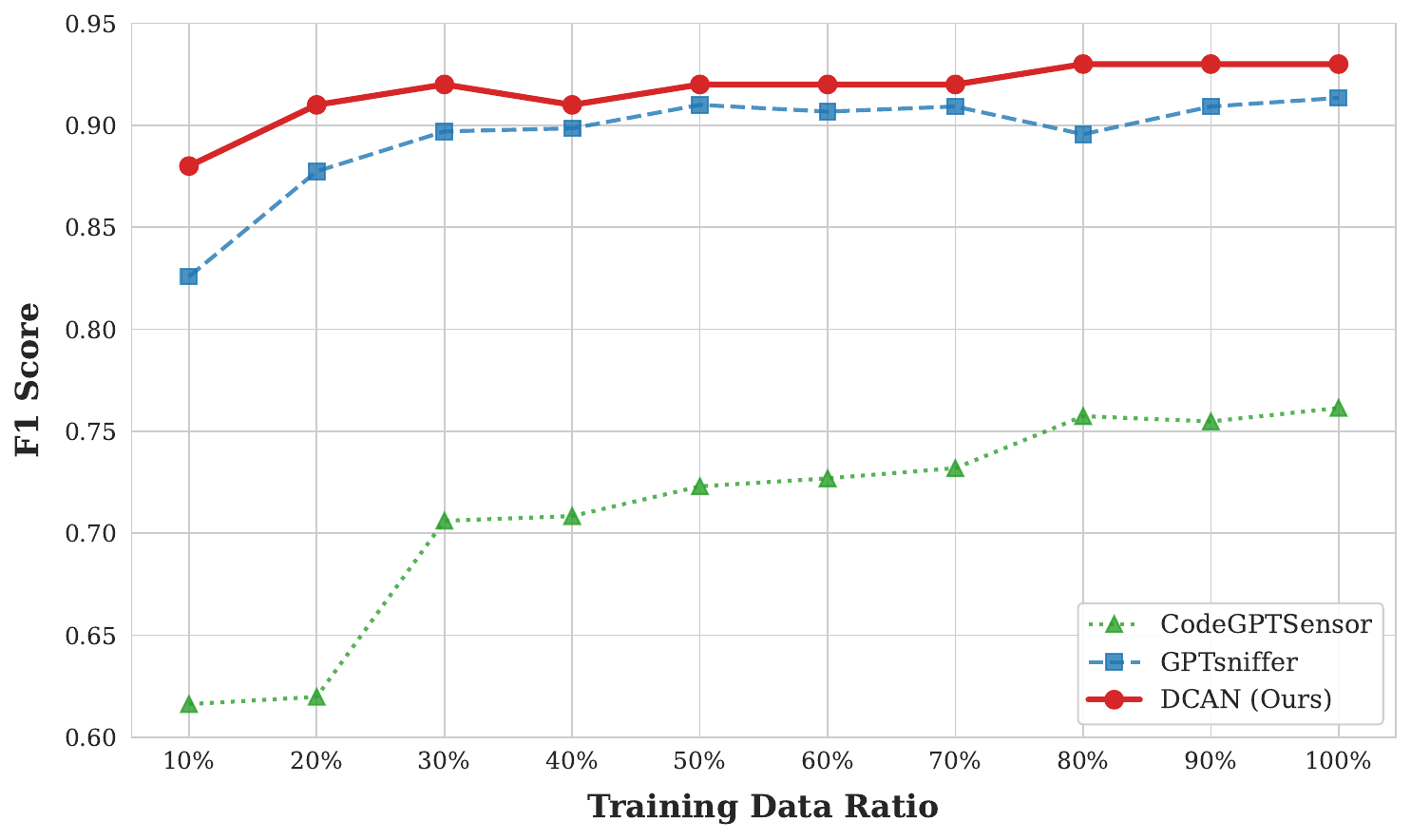}
    \vspace{-10pt}
    \caption{Performance comparison under varying training data ratios.}
    \vspace{-5pt}
    \label{fig:data_efficiency}
\end{figure}

We examine performance under different training data scales on the Java subdataset in the \textit{Plain} setting. The training data proportion is varied from 10\% to 100\% (in increments of 10\%), while the test set remains fixed. We compare DCAN with \textit{CodeGPTSensor} and \textit{GPTsniffer}. The results are shown in Figure~\ref{fig:data_efficiency}.
As observed in Figure~\ref{fig:data_efficiency}, DCAN consistently outperforms the baselines across all training scales. 
Even in the challenging setting with only \textbf{10\%} of the training data, DCAN achieves an F1-score of \textbf{88.03\%}, where \textit{CodeGPTSensor} achieves 61.62\%. This suggests that the disentanglement strategy enables DCAN to capture stable stylistic fingerprints without requiring large-scale training data.

As the training data increases, DCAN maintains a consistent advantage. While \textit{CodeGPTSensor} shows a gradual improvement (reaching 75.89\%), DCAN improves from 88.03\% to a peak of 93.20\%. These results demonstrate that DCAN achieves stable performance across varying data scales, which supports its practical applicability in data-constrained scenarios.

\subsubsection{Unified Generalist vs. Single-Language Specialists.}
In previous experiments, separate LLMCSA models were trained for each programming language. We next investigate whether a unified multilingual model can achieve comparable performance.

\begin{figure}[t]
    \centering
    \begin{subfigure}{0.48\columnwidth}
        \includegraphics[width=\linewidth]{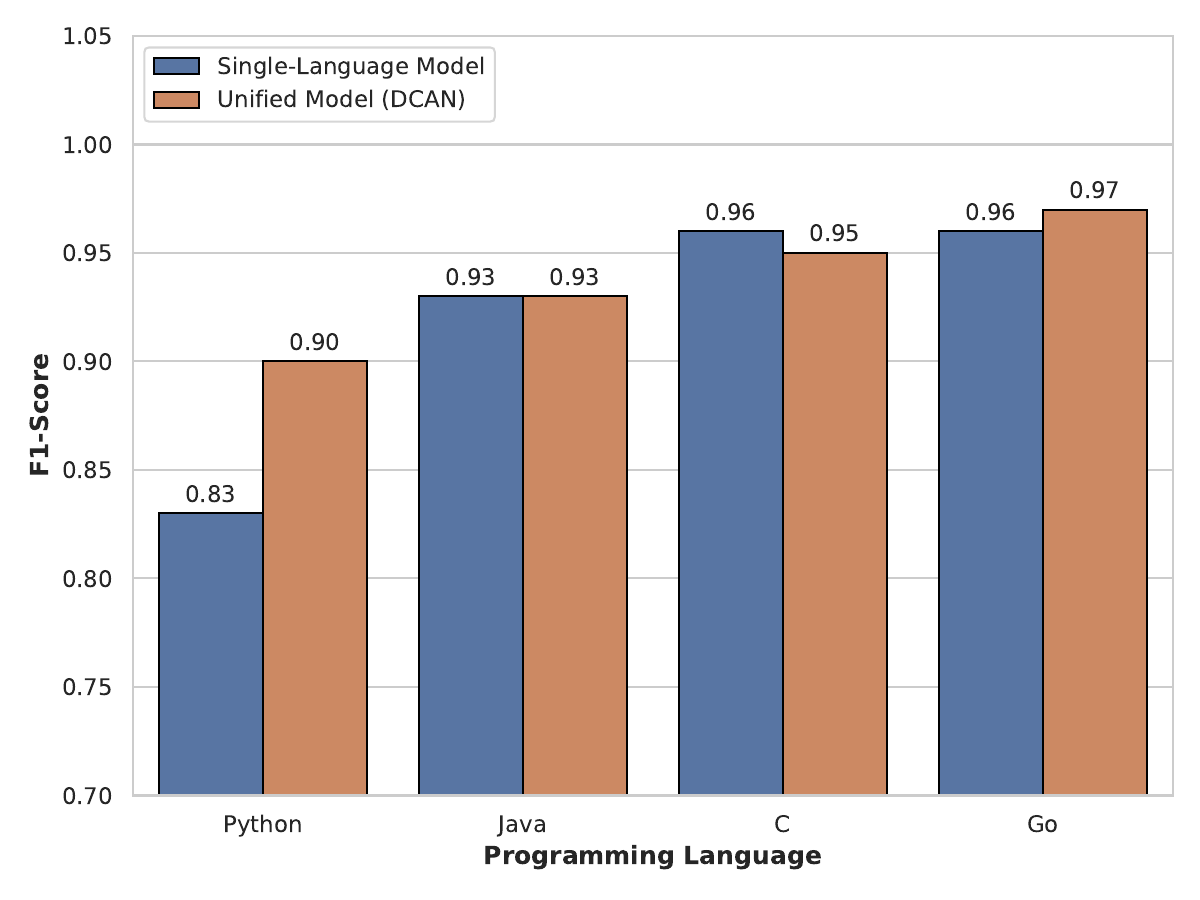}
        \caption{}
        \label{fig:lang_independence_plain}
    \end{subfigure}
    \hfill
    \begin{subfigure}{0.48\columnwidth}
        \includegraphics[width=\linewidth]{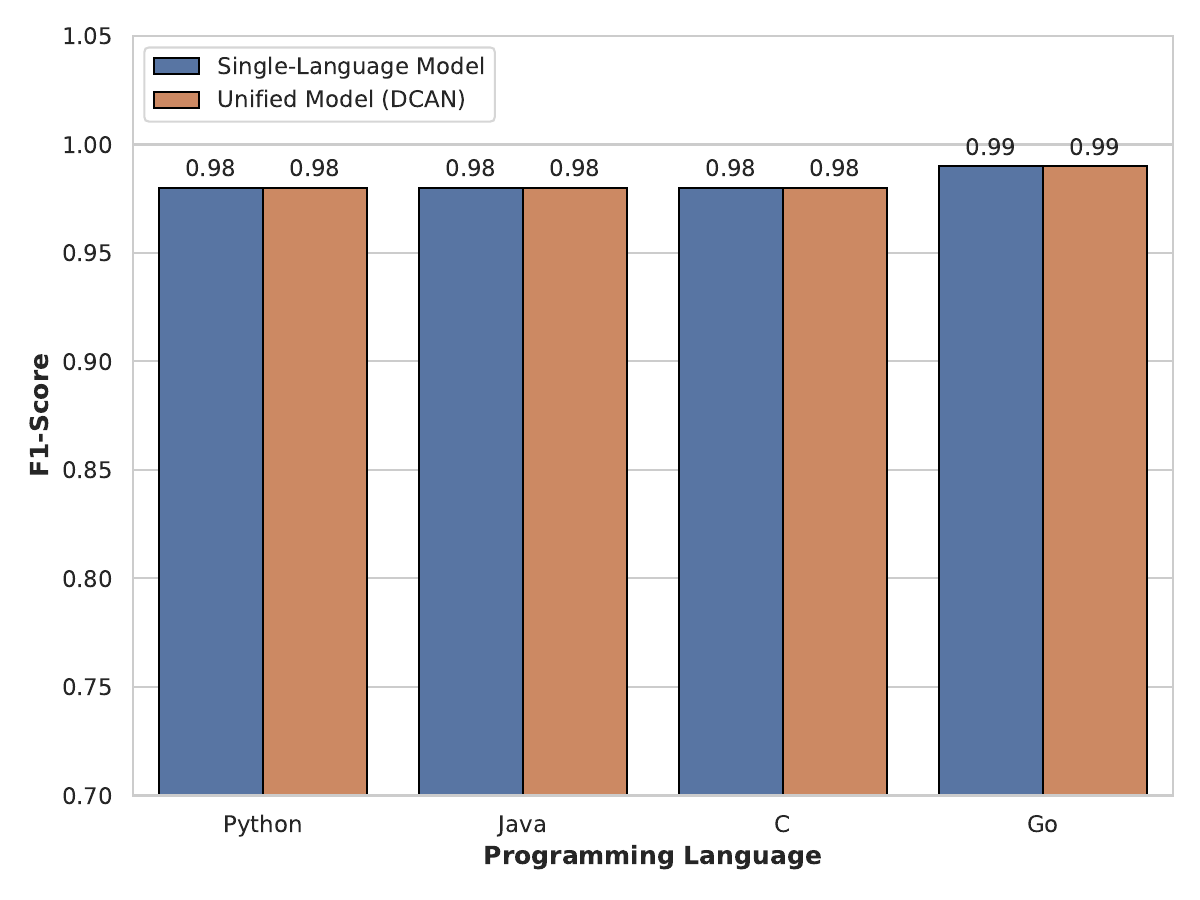}
        \caption{}
        \label{fig:lang_independence_comment}
    \end{subfigure}
    \vspace{-10pt}
    \caption{\textbf{Performance Comparison: Specialist vs. Generalist.} (a) and (b) denote the performance in plain and commented code, respectively.} 
    \vspace{-10pt}
    \label{fig:lang_independence_combined}
\end{figure}

As shown in Figure~\ref{fig:lang_independence_combined}, the unified model maintains performance comparable to language-specific models in Java and Go, with negligible degradation in C (less than 1\%). Notably, in Python, we observe a \textit{positive transfer} effect, where the F1-score improves from \textbf{0.83} to \textbf{0.90}. This improvement suggests that the Unified DCAN is able to capture model-dependent fingerprints that are partially transferable across programming languages.

\begin{figure*}[t]
    \centering
    \begin{subfigure}{0.24\textwidth}
        \includegraphics[width=\linewidth]{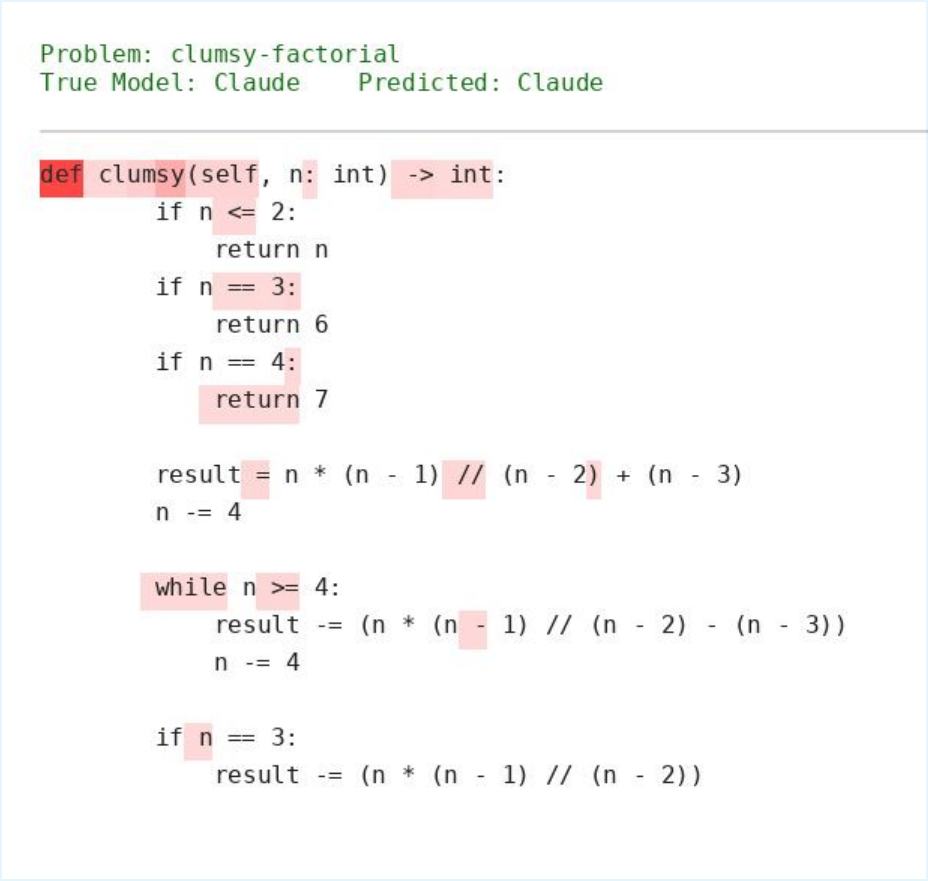}
        \caption{Claude}
    \end{subfigure}
    \hfill
    \begin{subfigure}{0.24\textwidth}
        \includegraphics[width=\linewidth]{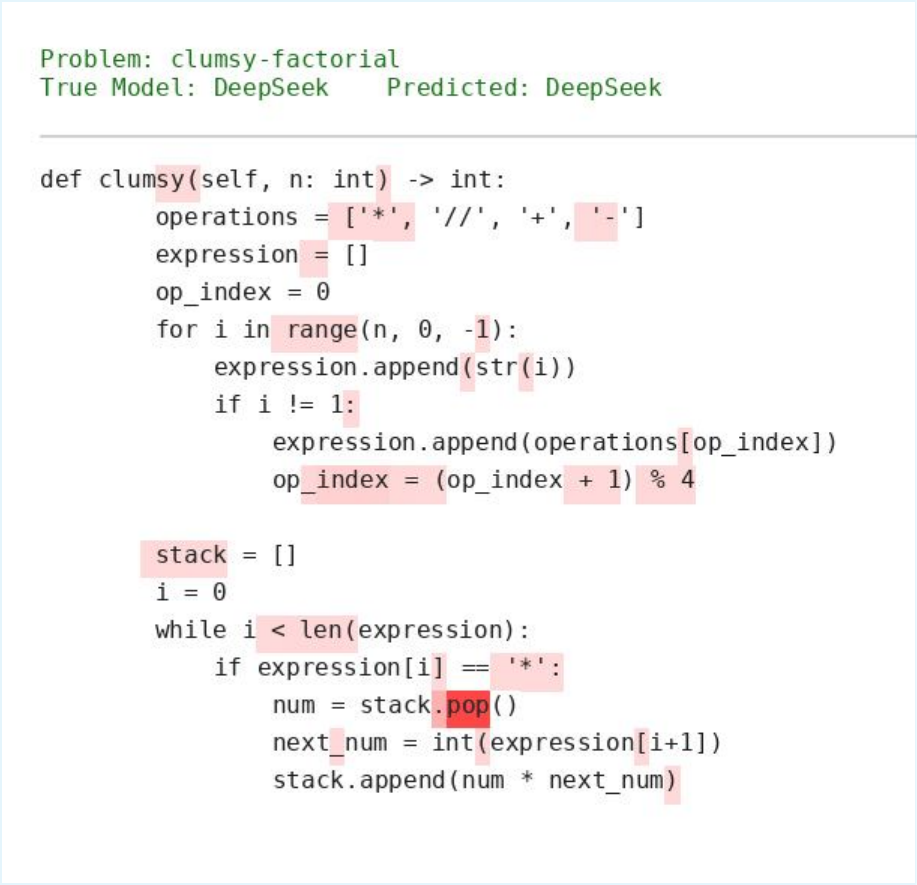}
        \caption{DeepSeek}
    \end{subfigure}
    \hfill
    \begin{subfigure}{0.24\textwidth}
        \includegraphics[width=\linewidth]{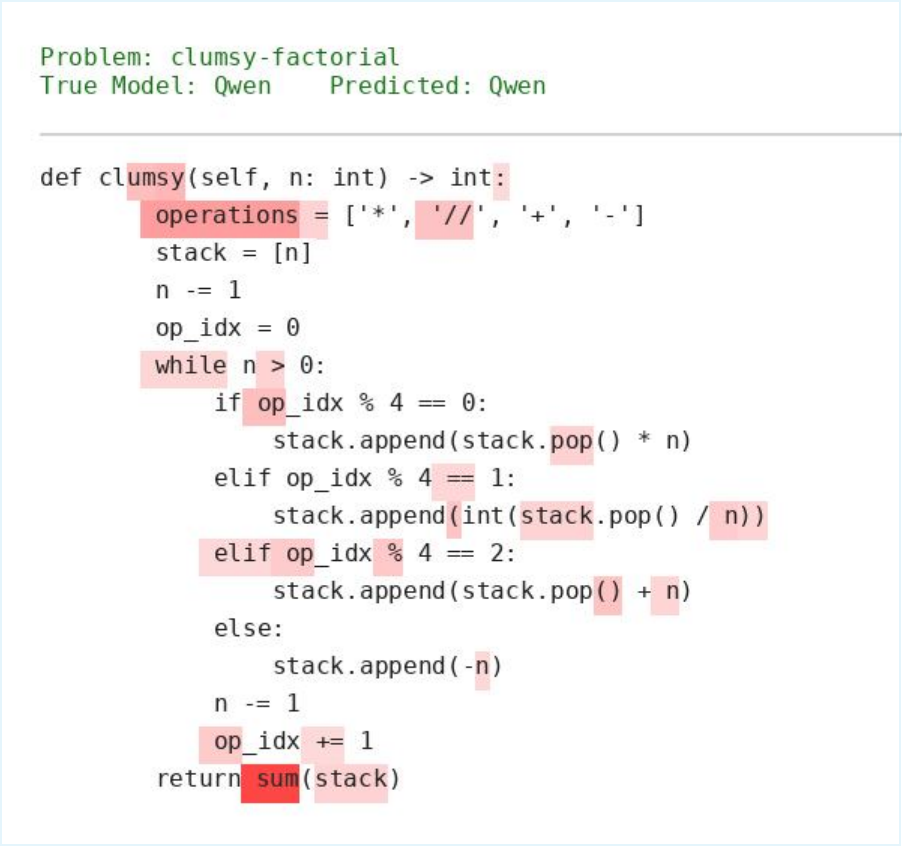}
        \caption{Qwen}
    \end{subfigure}
    \hfill
    \begin{subfigure}{0.24\textwidth}
        \includegraphics[width=\linewidth]{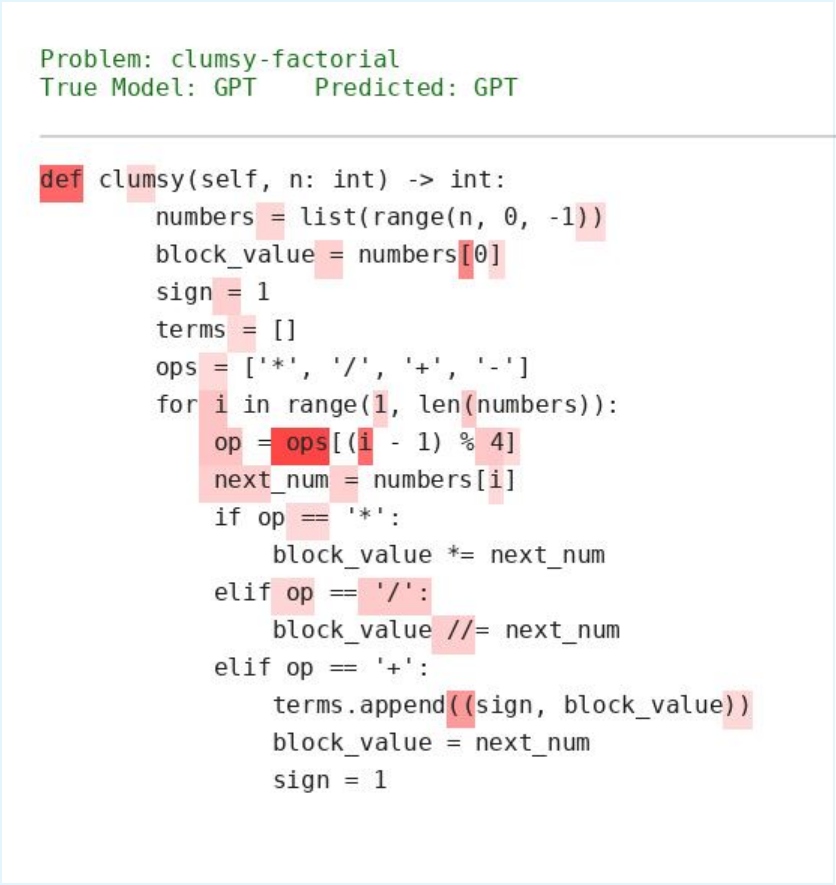}
        \caption{GPT}
    \end{subfigure}
    \caption{\textbf{Attribution Heatmaps on Plain Code.} The model identifies source LLMs based on structural features, including control flow constructs (Claude), stack-related APIs (DeepSeek), operator usage preferences (Qwen), and variable naming conventions~(GPT).}
    \label{fig:heatmap_plain}
\end{figure*}

\begin{figure*}[t]
    \centering
    \begin{subfigure}{0.24\textwidth}
        \includegraphics[width=\linewidth]{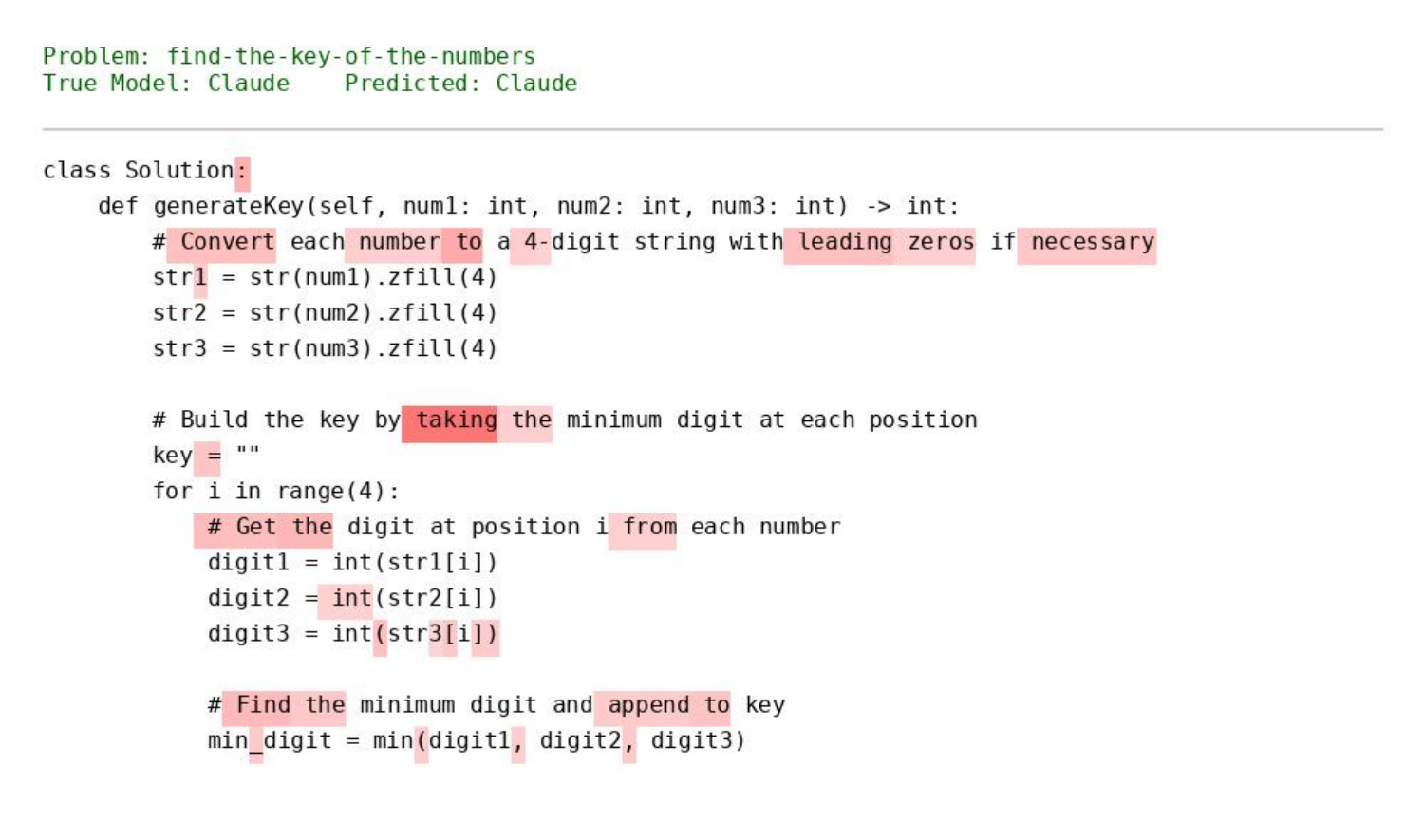}
        \caption{Claude}
    \end{subfigure}
    \hfill
    \begin{subfigure}{0.24\textwidth}
        \includegraphics[width=\linewidth]{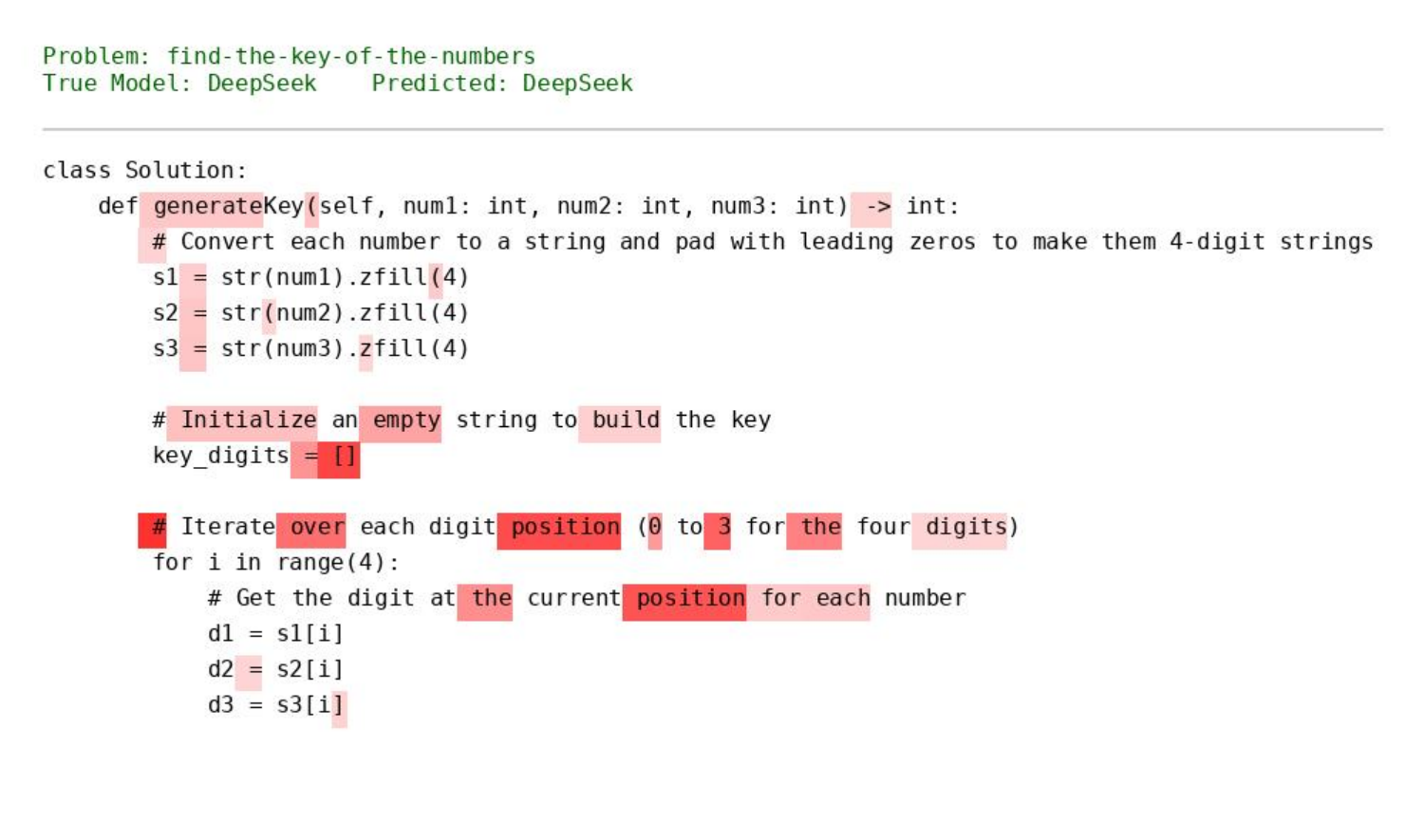}
        \caption{DeepSeek}
    \end{subfigure}
    \hfill
    \begin{subfigure}{0.24\textwidth}
        \includegraphics[width=\linewidth]{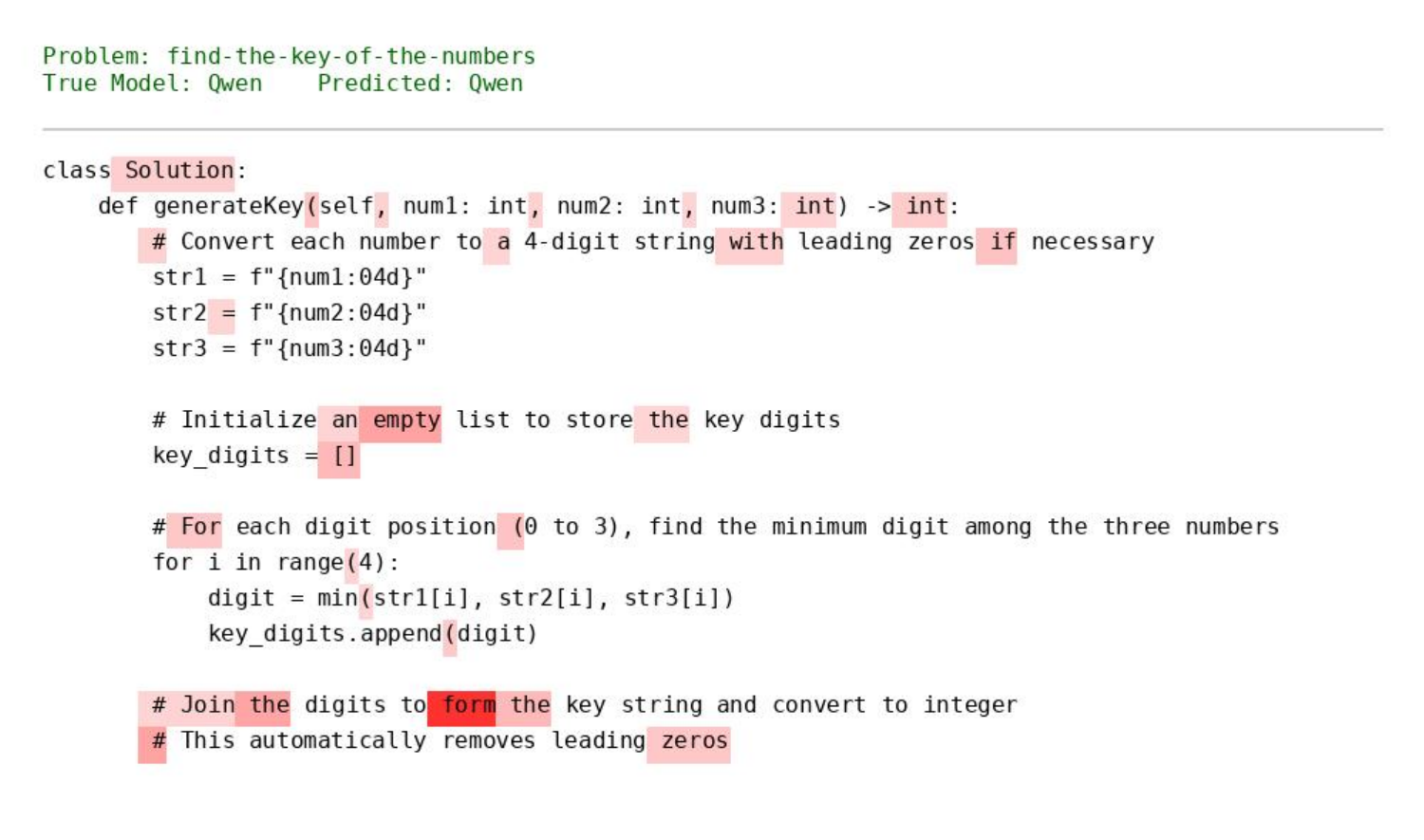}
        \caption{Qwen}
    \end{subfigure}
    \hfill
    \begin{subfigure}{0.24\textwidth}
        \includegraphics[width=\linewidth]{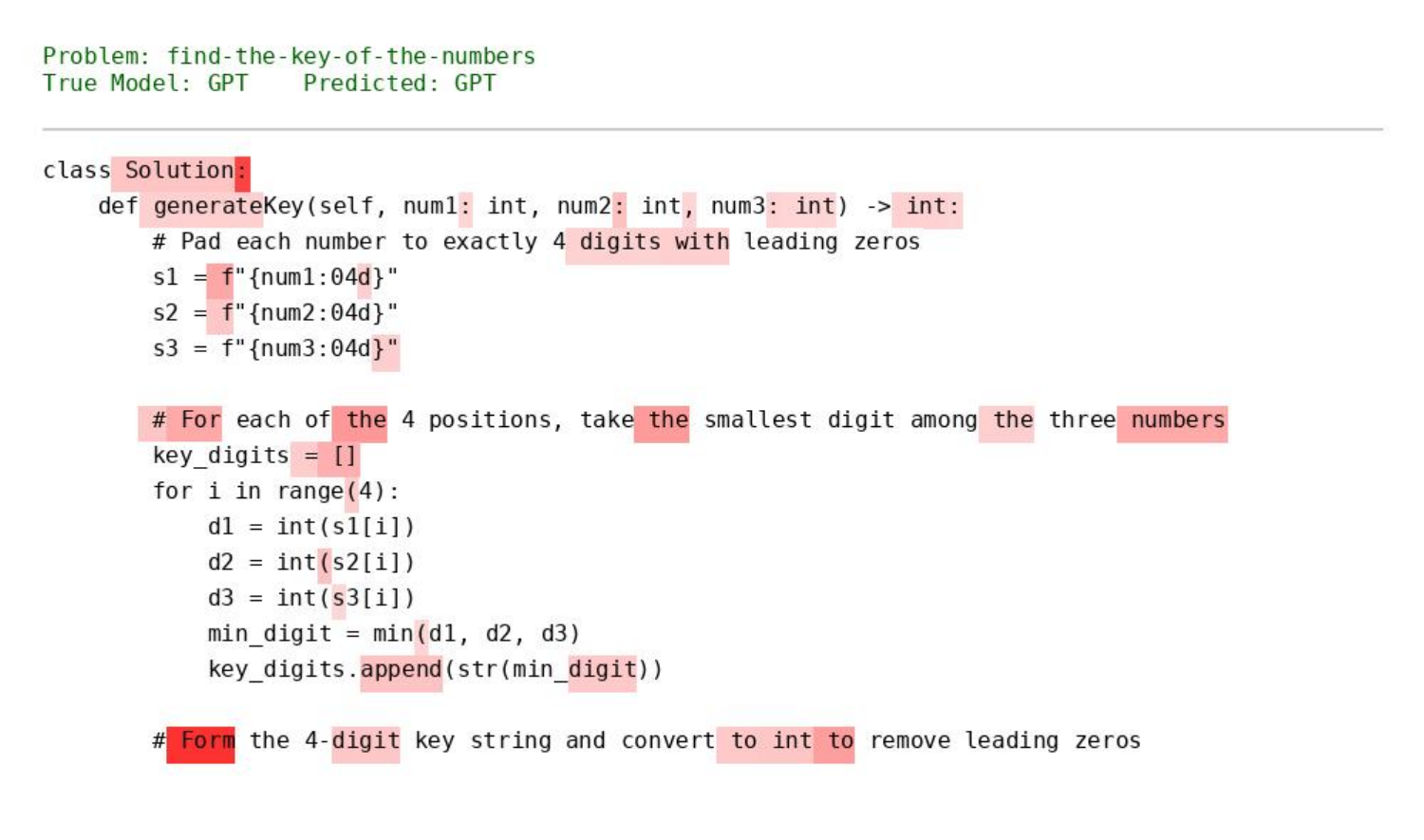}
        \caption{GPT}
    \end{subfigure}
    \caption{\textbf{Attribution Heatmaps on Commented Code.} In the presence of comments, the model utilizes a \textbf{dual-modality strategy}, combining specific syntactic idioms
    with auxiliary natural-language signals to support source attribution.}
    \label{fig:heatmap_comment}
\end{figure*}

\begin{table}[!t]
\centering
\caption{LOLO generalization results.}
\vspace{-5pt}
\label{tab:lodo_combined}
\begin{tabular}{lcccc}
\toprule
\textbf{Source Dataset} & \textbf{Target} & \textbf{Accuracy (\%)} & \textbf{F1-score} \\
\midrule
\multicolumn{4}{c}{\textit{Plain Setting}} \\
\midrule
Go+Java+Python & C      & 90.26 & 90.26 \\
C+Java+Python  & Go     & 83.94 & 83.94 \\
C+Go+Python    & Java   & 84.88 & 84.88 \\
C+Go+Java      & Python & 69.70 & 69.70 \\
\midrule
\multicolumn{4}{c}{\textit{Comment Setting}} \\
\midrule
Go+Java+Python & C      & 94.91 & 94.91 \\
C+Java+Python  & Go     & 96.72 & 96.72 \\
C+Go+Python    & Java   & 90.33 & 90.33 \\
C+Go+Java      & Python & 93.48 & 93.48 \\
\bottomrule
\end{tabular}
\vspace{-10pt}
\end{table}

\subsubsection{Zero-Shot Cross-Language Capability.}
To further assess generalization, we evaluate DCAN under a Leave-One-Language-Out (LOLO) setting. In this setup, the model is trained on three languages and directly evaluated on the fourth unseen language. The results are reported in Table~\ref{tab:lodo_combined}.

Under the \textit{Plain} setting, DCAN achieves strong zero-shot performance among syntactically similar languages (e.g., training on C/Java/Python and testing on Go). However, performance declines when generalizing to Python (69.70\%), likely due to the significant \textit{syntactic and structural differences} between Python and C-family languages. Unlike brace-based, statically typed languages, Python relies on indentation and dynamic constructs, leading to a distribution shift in structural representations.
In the \textit{Comment} setting, the inclusion of comments substantially improves cross-language generalization. Zero-shot accuracy on Python surges to 93.48\%, while Go reaches 96.72\%. These results indicate that, although programming syntax varies across languages, LLMs exhibit relatively consistent \textit{natural language generation style}, which DCAN leverages for robust, language-agnostic attribution.

\begin{findingbox}
\textbf{Finding 4:} DCAN learns transferable stylistic representations that support efficient training and cross-language generalization. The results indicate that the extracted fingerprints remain stable across varying data scales and programming languages.
\end{findingbox}

\subsection{Case Study}
\label{sec:case_study}

To qualitatively examine the stylistic fingerprints captured by DCAN, we apply gradient-based saliency analysis to two representative tasks. Figure~\ref{fig:heatmap_plain} visualizes the highlighted regions under the \textit{Plain} setting. Although the generated programs implement identical functionality, DCAN identifies discriminative regions associated with different source models. For instance, Claude-generated code places greater emphasis on explicit control-flow constructs (e.g., \texttt{if}/\texttt{elif}), DeepSeek frequently employs stack-related API operations (e.g., \texttt{.pop}), Qwen exhibits a preference for functional aggregation operations (e.g., \texttt{sum}), while GPT exhibits distinctive variable naming conventions (e.g., \texttt{next\_num}). These observations suggest that different LLMs exhibit consistent generation preferences that manifest as model-dependent fingerprints.

Under the \textit{Comment} setting (Figure~\ref{fig:heatmap_comment}), distinct coding idioms are observed across different models. For example, the \texttt{.zfill()} method is more frequently associated with Claude and DeepSeek, whereas f-string formatting (\texttt{f"\{...:04d\}"}) appears more commonly in GPT and Qwen. Beyond syntactic differences, variations in commenting style are also evident. DeepSeek tends to provide more descriptive annotations (e.g., ``Convert... and pad''), while GPT adopts comparatively concise expressions (e.g., ``Pad each number''). These findings indicate that comments provide complementary attribution signals, enabling joint exploitation of structural and linguistic information for more reliable source attribution.

\section{Conclusion}

In this paper, we investigate the LLMCSA task, which aims to attribute the source LLM responsible for generated code to support provenance analysis and accountability.
To facilitate systematic evaluation, we construct a large-scale LLMCSA benchmark dataset, which comprises code generated by four mainstream LLMs across four programming languages under two coding settings. The dataset provides a structured foundation for analyzing model-dependent fingerprints in LLM-generated code.
We further propose a disentanglement-based attribution framework that separates source-agnostic information from source-specific information. By explicitly modeling this decomposition, the framework improves the isolation of model-dependent fingerprints and enhances attribution performance.
Extensive experiments demonstrate that LLM-generated code exhibits consistent and distinguishable model-dependent fingerprints and that the proposed framework achieves reliable and robust source attribution across diverse settings.

\bibliographystyle{ACM-Reference-Format}
\bibliography{sample-base}


\begin{thebibliography}{28}


\ifx \showCODEN    \undefined \def \showCODEN     #1{\unskip}     \fi
\ifx \showISBNx    \undefined \def \showISBNx     #1{\unskip}     \fi
\ifx \showISBNxiii \undefined \def \showISBNxiii  #1{\unskip}     \fi
\ifx \showISSN     \undefined \def \showISSN      #1{\unskip}     \fi
\ifx \showLCCN     \undefined \def \showLCCN      #1{\unskip}     \fi
\ifx \shownote     \undefined \def \shownote      #1{#1}          \fi
\ifx \showarticletitle \undefined \def \showarticletitle #1{#1}   \fi
\ifx \showURL      \undefined \def \showURL       {\relax}        \fi
\providecommand\bibfield[2]{#2}
\providecommand\bibinfo[2]{#2}
\providecommand\natexlab[1]{#1}
\providecommand\showeprint[2][]{arXiv:#2}

\bibitem[Achiam et~al\mbox{.}(2023)]%
        {achiam2023gpt}
\bibfield{author}{\bibinfo{person}{Josh Achiam}, \bibinfo{person}{Steven Adler}, \bibinfo{person}{Sandhini Agarwal}, \bibinfo{person}{Lama Ahmad}, \bibinfo{person}{Ilge Akkaya}, \bibinfo{person}{Florencia~Leoni Aleman}, \bibinfo{person}{Diogo Almeida}, \bibinfo{person}{Janko Altenschmidt}, \bibinfo{person}{Sam Altman}, \bibinfo{person}{Shyamal Anadkat}, {et~al\mbox{.}}} \bibinfo{year}{2023}\natexlab{}.
\newblock \showarticletitle{Gpt-4 technical report}.
\newblock \bibinfo{journal}{\emph{arXiv preprint arXiv:2303.08774}} (\bibinfo{year}{2023}).
\newblock


\bibitem[Anthropic(2024)]%
        {anthropic2024claude3}
\bibfield{author}{\bibinfo{person}{Anthropic}.} \bibinfo{year}{2024}\natexlab{}.
\newblock \bibinfo{booktitle}{\emph{The Claude 3 Model Family: Opus, Sonnet, Haiku}}.
\newblock \bibinfo{type}{Model Card}. \bibinfo{institution}{Anthropic}.
\newblock
\urldef\tempurl%
\url{https://assets.anthropic.com/m/61e7d27f8c8f5919/original/Claude-3-Model-Card.pdf}
\showURL{%
\tempurl}


\bibitem[Bai et~al\mbox{.}(2023)]%
        {bai2023qwen}
\bibfield{author}{\bibinfo{person}{Jinze Bai}, \bibinfo{person}{Shuai Bai}, \bibinfo{person}{Yunfei Chu}, \bibinfo{person}{Zeyu Cui}, \bibinfo{person}{Kai Dang}, \bibinfo{person}{Xiaodong Deng}, \bibinfo{person}{Yang Fan}, \bibinfo{person}{Wenbin Ge}, \bibinfo{person}{Yu Han}, \bibinfo{person}{Fei Huang}, {et~al\mbox{.}}} \bibinfo{year}{2023}\natexlab{}.
\newblock \showarticletitle{Qwen technical report}.
\newblock \bibinfo{journal}{\emph{arXiv preprint arXiv:2309.16609}} (\bibinfo{year}{2023}).
\newblock


\bibitem[Bisztray et~al\mbox{.}(2025)]%
        {bisztray2025know}
\bibfield{author}{\bibinfo{person}{Tamas Bisztray}, \bibinfo{person}{Bilel Cherif}, \bibinfo{person}{Richard~A Dubniczky}, \bibinfo{person}{Nils Gruschka}, \bibinfo{person}{Bertalan Borsos}, \bibinfo{person}{Mohamed~Amine Ferrag}, \bibinfo{person}{Attila Kovacs}, \bibinfo{person}{Vasileios Mavroeidis}, {and} \bibinfo{person}{Norbert Tihanyi}.} \bibinfo{year}{2025}\natexlab{}.
\newblock \showarticletitle{I Know Which LLM Wrote Your Code Last Summer: LLM generated Code Stylometry for Authorship Attribution}. In \bibinfo{booktitle}{\emph{Proceedings of the 18th ACM Workshop on Artificial Intelligence and Security}}. \bibinfo{pages}{28--39}.
\newblock


\bibitem[Bulla et~al\mbox{.}(2024)]%
        {Bulla2024excode}
\bibfield{author}{\bibinfo{person}{Luana Bulla}, \bibinfo{person}{Alessandro Midolo}, \bibinfo{person}{Misael Mongiovì}, {and} \bibinfo{person}{Emiliano Tramontana}.} \bibinfo{year}{2024}\natexlab{}.
\newblock \showarticletitle{EX-CODE: A Robust and Explainable Model to Detect AI-Generated Code}.
\newblock \bibinfo{journal}{\emph{Information}} \bibinfo{volume}{15}, \bibinfo{number}{12} (\bibinfo{year}{2024}).
\newblock
\showISSN{2078-2489}
\href{https://doi.org/10.3390/info15120819}{doi:\nolinkurl{10.3390/info15120819}}


\bibitem[Dubey et~al\mbox{.}(2024)]%
        {dubey2024llama}
\bibfield{author}{\bibinfo{person}{Abhimanyu Dubey}, \bibinfo{person}{Abhinav Jauhri}, \bibinfo{person}{Abhinav Pandey}, \bibinfo{person}{Abhishek Kadian}, \bibinfo{person}{Ahmad Al-Dahle}, \bibinfo{person}{Aiesha Letman}, \bibinfo{person}{Akhil Mathur}, \bibinfo{person}{Alan Schelten}, \bibinfo{person}{Amy Yang}, \bibinfo{person}{Angela Fan}, {et~al\mbox{.}}} \bibinfo{year}{2024}\natexlab{}.
\newblock \showarticletitle{The llama 3 herd of models}.
\newblock \bibinfo{journal}{\emph{arXiv e-prints}} (\bibinfo{year}{2024}), \bibinfo{pages}{arXiv--2407}.
\newblock


\bibitem[Feng(2020)]%
        {feng2020codebert}
\bibfield{author}{\bibinfo{person}{Z Feng}.} \bibinfo{year}{2020}\natexlab{}.
\newblock \showarticletitle{Codebert: A pre-trained model for program-ming and natural languages}.
\newblock \bibinfo{journal}{\emph{arXiv preprint arXiv:2002.08155}} (\bibinfo{year}{2020}).
\newblock


\bibitem[Guo et~al\mbox{.}(2022)]%
        {guo2022unixcoder}
\bibfield{author}{\bibinfo{person}{Daya Guo}, \bibinfo{person}{Shuai Lu}, \bibinfo{person}{Nan Duan}, \bibinfo{person}{Yanlin Wang}, \bibinfo{person}{Ming Zhou}, {and} \bibinfo{person}{Jian Yin}.} \bibinfo{year}{2022}\natexlab{}.
\newblock \showarticletitle{Unixcoder: Unified cross-modal pre-training for code representation}.
\newblock \bibinfo{journal}{\emph{arXiv preprint arXiv:2203.03850}} (\bibinfo{year}{2022}).
\newblock


\bibitem[Huang et~al\mbox{.}(2024)]%
        {huang2024can}
\bibfield{author}{\bibinfo{person}{Baixiang Huang}, \bibinfo{person}{Canyu Chen}, {and} \bibinfo{person}{Kai Shu}.} \bibinfo{year}{2024}\natexlab{}.
\newblock \showarticletitle{Can large language models identify authorship?}
\newblock \bibinfo{journal}{\emph{arXiv preprint arXiv:2403.08213}} (\bibinfo{year}{2024}).
\newblock


\bibitem[Jiang et~al\mbox{.}(2024)]%
        {jiang2024survey}
\bibfield{author}{\bibinfo{person}{Juyong Jiang}, \bibinfo{person}{Fan Wang}, \bibinfo{person}{Jiasi Shen}, \bibinfo{person}{Sungju Kim}, {and} \bibinfo{person}{Sunghun Kim}.} \bibinfo{year}{2024}\natexlab{}.
\newblock \showarticletitle{A survey on large language models for code generation}.
\newblock \bibinfo{journal}{\emph{ACM Transactions on Software Engineering and Methodology}} (\bibinfo{year}{2024}).
\newblock


\bibitem[Jimenez et~al\mbox{.}(2023)]%
        {jimenez2023swe}
\bibfield{author}{\bibinfo{person}{Carlos~E Jimenez}, \bibinfo{person}{John Yang}, \bibinfo{person}{Alexander Wettig}, \bibinfo{person}{Shunyu Yao}, \bibinfo{person}{Kexin Pei}, \bibinfo{person}{Ofir Press}, {and} \bibinfo{person}{Karthik Narasimhan}.} \bibinfo{year}{2023}\natexlab{}.
\newblock \showarticletitle{Swe-bench: Can language models resolve real-world github issues?}
\newblock \bibinfo{journal}{\emph{arXiv preprint arXiv:2310.06770}} (\bibinfo{year}{2023}).
\newblock


\bibitem[Koike et~al\mbox{.}(2024)]%
        {koike2024outfox}
\bibfield{author}{\bibinfo{person}{Ryuto Koike}, \bibinfo{person}{Masahiro Kaneko}, {and} \bibinfo{person}{Naoaki Okazaki}.} \bibinfo{year}{2024}\natexlab{}.
\newblock \showarticletitle{Outfox: Llm-generated essay detection through in-context learning with adversarially generated examples}. In \bibinfo{booktitle}{\emph{Proceedings of the AAAI Conference on Artificial Intelligence}}, Vol.~\bibinfo{volume}{38}. \bibinfo{pages}{21258--21266}.
\newblock


\bibitem[Liu et~al\mbox{.}(2024a)]%
        {liu2024deepseekv2}
\bibfield{author}{\bibinfo{person}{Aixin Liu}, \bibinfo{person}{Bei Feng}, \bibinfo{person}{Bin Wang}, \bibinfo{person}{Bingxuan Wang}, \bibinfo{person}{Bo Liu}, \bibinfo{person}{Chenggang Zhao}, \bibinfo{person}{Chengqi Dengr}, \bibinfo{person}{Chong Ruan}, \bibinfo{person}{Damai Dai}, \bibinfo{person}{Daya Guo}, {et~al\mbox{.}}} \bibinfo{year}{2024}\natexlab{a}.
\newblock \showarticletitle{Deepseek-v2: A strong, economical, and efficient mixture-of-experts language model}.
\newblock \bibinfo{journal}{\emph{arXiv preprint arXiv:2405.04434}} (\bibinfo{year}{2024}).
\newblock


\bibitem[Liu et~al\mbox{.}(2024b)]%
        {liu2024deepseek}
\bibfield{author}{\bibinfo{person}{Aixin Liu}, \bibinfo{person}{Bei Feng}, \bibinfo{person}{Bing Xue}, \bibinfo{person}{Bingxuan Wang}, \bibinfo{person}{Bochao Wu}, \bibinfo{person}{Chengda Lu}, \bibinfo{person}{Chenggang Zhao}, \bibinfo{person}{Chengqi Deng}, \bibinfo{person}{Chenyu Zhang}, \bibinfo{person}{Chong Ruan}, {et~al\mbox{.}}} \bibinfo{year}{2024}\natexlab{b}.
\newblock \showarticletitle{Deepseek-v3 technical report}.
\newblock \bibinfo{journal}{\emph{arXiv preprint arXiv:2412.19437}} (\bibinfo{year}{2024}).
\newblock


\bibitem[Lu et~al\mbox{.}(2025)]%
        {lu2025source}
\bibfield{author}{\bibinfo{person}{Xinyang Lu}, \bibinfo{person}{Jingtan Wang}, \bibinfo{person}{Zitong Zhao}, \bibinfo{person}{Zhongxiang Dai}, \bibinfo{person}{Chuan-Sheng Foo}, \bibinfo{person}{See-Kiong Ng}, {and} \bibinfo{person}{Bryan Kian~Hsiang Low}.} \bibinfo{year}{2025}\natexlab{}.
\newblock \bibinfo{title}{Source Attribution for Large Language Model-Generated Data}.
\newblock
\urldef\tempurl%
\url{https://openreview.net/forum?id=1ou5noWgHM}
\showURL{%
\tempurl}


\bibitem[Mitchell et~al\mbox{.}(2023)]%
        {mitchell2023detectgpt}
\bibfield{author}{\bibinfo{person}{Eric Mitchell}, \bibinfo{person}{Yoonho Lee}, \bibinfo{person}{Alexander Khazatsky}, \bibinfo{person}{Christopher~D Manning}, {and} \bibinfo{person}{Chelsea Finn}.} \bibinfo{year}{2023}\natexlab{}.
\newblock \showarticletitle{Detectgpt: Zero-shot machine-generated text detection using probability curvature}. In \bibinfo{booktitle}{\emph{International conference on machine learning}}. PMLR, \bibinfo{pages}{24950--24962}.
\newblock


\bibitem[Nguyen et~al\mbox{.}(2024)]%
        {nguyen2024gptsniffer}
\bibfield{author}{\bibinfo{person}{Phuong~T Nguyen}, \bibinfo{person}{Juri Di~Rocco}, \bibinfo{person}{Claudio Di~Sipio}, \bibinfo{person}{Riccardo Rubei}, \bibinfo{person}{Davide Di~Ruscio}, {and} \bibinfo{person}{Massimiliano Di~Penta}.} \bibinfo{year}{2024}\natexlab{}.
\newblock \showarticletitle{GPTSniffer: A CodeBERT-based classifier to detect source code written by ChatGPT}.
\newblock \bibinfo{journal}{\emph{Journal of Systems and Software}}  \bibinfo{volume}{214} (\bibinfo{year}{2024}), \bibinfo{pages}{112059}.
\newblock


\bibitem[Orel et~al\mbox{.}(2025)]%
        {orel2025codet}
\bibfield{author}{\bibinfo{person}{Daniil Orel}, \bibinfo{person}{Dilshod Azizov}, {and} \bibinfo{person}{Preslav Nakov}.} \bibinfo{year}{2025}\natexlab{}.
\newblock \showarticletitle{CoDet-M4: Detecting Machine-Generated Code in Multi-Lingual, Multi-Generator and Multi-Domain Settings}.
\newblock \bibinfo{journal}{\emph{arXiv preprint arXiv:2503.13733}} (\bibinfo{year}{2025}).
\newblock


\bibitem[Pan et~al\mbox{.}(2024)]%
        {pan2024assessing}
\bibfield{author}{\bibinfo{person}{Wei~Hung Pan}, \bibinfo{person}{Ming~Jie Chok}, \bibinfo{person}{Jonathan Leong~Shan Wong}, \bibinfo{person}{Yung~Xin Shin}, \bibinfo{person}{Yeong~Shian Poon}, \bibinfo{person}{Zhou Yang}, \bibinfo{person}{Chun~Yong Chong}, \bibinfo{person}{David Lo}, {and} \bibinfo{person}{Mei~Kuan Lim}.} \bibinfo{year}{2024}\natexlab{}.
\newblock \showarticletitle{Assessing ai detectors in identifying ai-generated code: Implications for education}. In \bibinfo{booktitle}{\emph{Proceedings of the 46th international conference on software engineering: software engineering education and training}}. \bibinfo{pages}{1--11}.
\newblock


\bibitem[Poldrack et~al\mbox{.}(2023)]%
        {poldrack2023ai}
\bibfield{author}{\bibinfo{person}{Russell~A Poldrack}, \bibinfo{person}{Thomas Lu}, {and} \bibinfo{person}{Ga{\v{s}}per Begu{\v{s}}}.} \bibinfo{year}{2023}\natexlab{}.
\newblock \showarticletitle{AI-assisted coding: Experiments with GPT-4}.
\newblock \bibinfo{journal}{\emph{arXiv preprint arXiv:2304.13187}} (\bibinfo{year}{2023}).
\newblock


\bibitem[Suresh et~al\mbox{.}(2024)]%
        {suresh2024watermarking}
\bibfield{author}{\bibinfo{person}{Tarun Suresh}, \bibinfo{person}{Shubham Ugare}, \bibinfo{person}{Gagandeep Singh}, {and} \bibinfo{person}{Sasa Misailovic}.} \bibinfo{year}{2024}\natexlab{}.
\newblock \showarticletitle{Is The Watermarking Of LLM-Generated Code Robust?}
\newblock \bibinfo{journal}{\emph{arXiv preprint arXiv:2403.17983}} (\bibinfo{year}{2024}).
\newblock


\bibitem[Tihanyi et~al\mbox{.}(2025)]%
        {tihanyi2025hidden}
\bibfield{author}{\bibinfo{person}{Norbert Tihanyi}, \bibinfo{person}{Bilel Cherif}, \bibinfo{person}{Richard~A Dubniczky}, \bibinfo{person}{Mohamed~Amine Ferrag}, {and} \bibinfo{person}{Tam{\'a}s Bisztray}.} \bibinfo{year}{2025}\natexlab{}.
\newblock \showarticletitle{The Hidden DNA of LLM-Generated JavaScript: Structural Patterns Enable High-Accuracy Authorship Attribution}.
\newblock \bibinfo{journal}{\emph{arXiv preprint arXiv:2510.10493}} (\bibinfo{year}{2025}).
\newblock


\bibitem[Wu et~al\mbox{.}(2025)]%
        {wu2025wrote}
\bibfield{author}{\bibinfo{person}{Junchao Wu}, \bibinfo{person}{Runzhe Zhan}, \bibinfo{person}{Derek~F Wong}, \bibinfo{person}{Shu Yang}, \bibinfo{person}{Xuebo Liu}, \bibinfo{person}{Lidia~S Chao}, {and} \bibinfo{person}{Min Zhang}.} \bibinfo{year}{2025}\natexlab{}.
\newblock \showarticletitle{Who wrote this? the key to zero-shot llm-generated text detection is gecscore}. In \bibinfo{booktitle}{\emph{Proceedings of the 31st International Conference on Computational Linguistics}}. \bibinfo{pages}{10275--10292}.
\newblock


\bibitem[Xia et~al\mbox{.}(2025)]%
        {xia2025leetcodedataset}
\bibfield{author}{\bibinfo{person}{Yunhui Xia}, \bibinfo{person}{Wei Shen}, \bibinfo{person}{Yan Wang}, \bibinfo{person}{Jason~Klein Liu}, \bibinfo{person}{Huifeng Sun}, \bibinfo{person}{Siyue Wu}, \bibinfo{person}{Jian Hu}, {and} \bibinfo{person}{Xiaolong Xu}.} \bibinfo{year}{2025}\natexlab{}.
\newblock \showarticletitle{Leetcodedataset: A temporal dataset for robust evaluation and efficient training of code llms}.
\newblock \bibinfo{journal}{\emph{arXiv preprint arXiv:2504.14655}} (\bibinfo{year}{2025}).
\newblock


\bibitem[Xu et~al\mbox{.}(2025)]%
        {xu2025distinguishing}
\bibfield{author}{\bibinfo{person}{Xiaodan Xu}, \bibinfo{person}{Chao Ni}, \bibinfo{person}{Xinrong Guo}, \bibinfo{person}{Shaoxuan Liu}, \bibinfo{person}{Xiaoya Wang}, \bibinfo{person}{Kui Liu}, {and} \bibinfo{person}{Xiaohu Yang}.} \bibinfo{year}{2025}\natexlab{}.
\newblock \showarticletitle{Distinguishing llm-generated from human-written code by contrastive learning}.
\newblock \bibinfo{journal}{\emph{ACM Transactions on Software Engineering and Methodology}} \bibinfo{volume}{34}, \bibinfo{number}{4} (\bibinfo{year}{2025}), \bibinfo{pages}{1--31}.
\newblock


\bibitem[Xu and Sheng(2024)]%
        {xu2024detecting}
\bibfield{author}{\bibinfo{person}{Zhenyu Xu} {and} \bibinfo{person}{Victor~S Sheng}.} \bibinfo{year}{2024}\natexlab{}.
\newblock \showarticletitle{Detecting AI-generated code assignments using perplexity of large language models}. In \bibinfo{booktitle}{\emph{Proceedings of the aaai conference on artificial intelligence}}, Vol.~\bibinfo{volume}{38}. \bibinfo{pages}{23155--23162}.
\newblock


\bibitem[Yang et~al\mbox{.}(2023)]%
        {yang2023zero}
\bibfield{author}{\bibinfo{person}{Xianjun Yang}, \bibinfo{person}{Kexun Zhang}, \bibinfo{person}{Haifeng Chen}, \bibinfo{person}{Linda Petzold}, \bibinfo{person}{William~Yang Wang}, {and} \bibinfo{person}{Wei Cheng}.} \bibinfo{year}{2023}\natexlab{}.
\newblock \showarticletitle{Zero-shot detection of machine-generated codes}.
\newblock \bibinfo{journal}{\emph{arXiv preprint arXiv:2310.05103}} (\bibinfo{year}{2023}).
\newblock


\bibitem[Ye et~al\mbox{.}(2025)]%
        {ye2025uncovering}
\bibfield{author}{\bibinfo{person}{Tong Ye}, \bibinfo{person}{Yangkai Du}, \bibinfo{person}{Tengfei Ma}, \bibinfo{person}{Lingfei Wu}, \bibinfo{person}{Xuhong Zhang}, \bibinfo{person}{Shouling Ji}, {and} \bibinfo{person}{Wenhai Wang}.} \bibinfo{year}{2025}\natexlab{}.
\newblock \showarticletitle{Uncovering llm-generated code: A zero-shot synthetic code detector via code rewriting}. In \bibinfo{booktitle}{\emph{Proceedings of the AAAI Conference on Artificial Intelligence}}, Vol.~\bibinfo{volume}{39}. \bibinfo{pages}{968--976}.
\newblock


\end{thebibliography}

\end{document}